\documentclass[iop]{emulateapj}

\usepackage{xcolor}
\usepackage[colorlinks]{hyperref}
\hypersetup{
    colorlinks = true,
    citecolor = {blue}
}

\newcommand{\cii}{C~{\scriptsize II}}
\newcommand{\scii}{C~{\tiny II}}

\slugcomment{Not to appear in Nonlearned J., 45.}

\shorttitle{SEDs of Companion galaxies to z$\sim$6 QSOs}
\shortauthors{Mazzucchelli et al.}

\begin{document}

\title{Spectral Energy Distributions of Companion Galaxies to z$\sim$6 Quasars}

\author{Mazzucchelli C.\altaffilmark{1,2, $\dagger$},
Decarli R.\altaffilmark{3},
Farina E. P.\altaffilmark{4,1},
Ba\~nados E.\altaffilmark{1,5},
Venemans B. P.\altaffilmark{1},
Strauss M. A.\altaffilmark{6},
Walter F.\altaffilmark{1,7,8},
Neeleman M.\altaffilmark{1},
Bertoldi F.\altaffilmark{9},
Fan X.\altaffilmark{10},
Riechers D.\altaffilmark{11},
Rix H.-W.\altaffilmark{1},
and
Wang R.\altaffilmark{12}
}
\affil{\altaffilmark{1}Max-Planck-Institut f$\rm \ddot{u}$r Astronomie, K{\"o}nigstuhl 17, D-69117 Heidelberg, Germany}
\affil{\altaffilmark{2}European Southern Observatory, Alonso de Cordova 3107, Vitacura, Region Metropolitana, Chile}
\affil{\altaffilmark{3}INAF--Osservatorio di Astrofisica e Scienza dello Spazio, via Gobetti 93/3, I-40129, Bologna, Italy}
\affil{\altaffilmark{4}Department of Physics, Broida Hall, University of California, Santa Barbara, CA 93106–9530, USA}
\affil{\altaffilmark{5}The Observatories of the Carnegie Institution for Science, 813 Santa Barbara Street, Pasadena, CA 91101, USA}
\affil{\altaffilmark{6}Department of Astrophysical Sciences, Princeton University, Princeton, NJ 08544, USA}
\affil{\altaffilmark{7}National Radio Astronomy Observatory, Pete V.~Domenici Array Institute Science Center, P.O.~Box O, Socorro, NM 87801, USA}
\affil{\altaffilmark{8}Astronomy Department, California Institute of Technology, MC249-17, Pasadena, CA 91125, USA}
\affil{\altaffilmark{9}Argelander Institute for Astronomy, University of Bonn, Auf dem H\"ugel 71, D-53121 Bonn, Germany}
\affil{\altaffilmark{10}Steward Observatory, University of Arizona, 933 N.~Cherry Street, Tucson, AZ 85721, USA}
\affil{\altaffilmark{11}Cornell University, 220 Space Sciences Building, Ithaca, NY 14853, USA}
\affil{\altaffilmark{12}Kavli Institute of Astronomy and Astrophysics at Peking University, 5 Yiheyuan Road, Haidian District, Beijing 100871, People's Republic of China}
\affil{\altaffilmark{$\dagger$}ESO Fellow}

\begin{abstract}
Massive, quiescent galaxies are already observed at redshift $z\sim4$, i.e.~$\sim$1.5 Gyr after the Big Bang.
Current models predict them to be formed via massive, gas--rich mergers at $z>6$.
Recent ALMA observations of the cool gas and dust in $z\gtrsim$6 quasars have discovered [\cii]-- and far infrared--bright galaxies adjacent to several quasars.
In this work, we present sensitive imaging and spectroscopic follow-up observations, with \textit{HST}/WFC3, \textit{Spitzer}/IRAC, VLT/MUSE, Magellan/FIRE and LBT/LUCI-MODS, of ALMA-detected, dust-rich companion galaxies of four quasars at $z\gtrsim 6$, specifically acquired to probe their stellar content and unobscured star formation rate.
Three companion galaxies do not show significant emission in the observed optical/IR wavelength range.
The photometric limits suggest that these galaxies are highly dust--enshrouded, with unobscured star formation rates SFR$_{\rm UV}<$few M$_{\odot}$ yr$^{-1}$, and a stellar content of $M_{*}<$10$^{10}$ M$_{\odot}$ yr$^{-1}$. 
However, the companion to PJ167-13 shows bright rest--frame UV emission (F140W AB = 25.48).
Its SED resembles that of a star--forming galaxy with a total SFR$\sim$50 M$_{\odot}$ yr$^{-1}$ and $M_{*}\sim 9 \times 10^{9}$ M$_{\odot}$.
All the companion sources are consistent with residing on the galaxy main sequence at $z\sim$6.
Additional, deeper data from future facilities, such as \textit{JWST}, are needed in order to characterize these gas--rich sources in the first Gyr of cosmic history.
\end{abstract}


\section{Introduction} \label{secIntro}
A large population of massive ($\sim10^{11}$ M$_{\odot}$), quiescent and compact galaxies has been observed at very early cosmic times ($2 < z < 4$) when the universe was between $\sim$1.6 and $\sim$3.6 Gyr old (e.g.~\citealt{vanDokkum08}, \citealt{Straatman14}).
Several studies suggest that these galaxies formed from gas-rich, major mergers at 3$\lesssim z\lesssim$7 (e.g.~\citealt{Wuyts10}, \citealt{Hopkins08}).
High-redshift ($z\gtrsim$3) sub-millimeter galaxies (SMGs) have been invoked as possible progenitors of these massive $z\gtrsim$2 ``red-and-dead'' sources (e.g.~\citealt{Toft14}).

SMGs are galaxies with large infrared luminosities ($L_{\mathrm{IR}}\gtrsim$10$^{12}$ M$_{\odot}$), which are thought to experience short and intense episodes of star formation (e.g.~\citealt{Blain02}, \citealt{Spilker14}). 
Recent studies show that SMGs at $z\sim$3.5 often have sub-components or companions when observed at kpc scale resolution (e.g.~\citealt{Hodge13a}, \citealt{Gomez18}), or are located in dense environments (e.g.~\citealt{Hodge13b}, \citealt{Riechers14}), suggesting that these sources have experienced recent mergers.
At even higher redshifts ($z\gtrsim$6), only a few SMGs not hosting a central active galactic nucleus have been observed.
\cite{Riechers13} found a dust-obscured, extremely star--forming SMG at $z=$6.3 with a star formation rate SFR $\sim$3000 M$_{\odot}$ yr$^{-1}$; \cite{Fudamoto17} and \cite{Zavala18} discovered a $z=$6.03 galaxy whose SFR was somewhat smaller ($\sim$950 M$_{\odot}$ yr$^{-1}$); \cite{Marrone18} discovered a pair of massive SMGs with similar SFR at $z\sim$6.9.
The only other highly star--forming ($\gtrsim$100 M$_{\odot}$ yr$^{-1}$) objects known thus far at such redshifts, with luminosities extending to slightly fainter values ($L_{\rm IR}\sim$few 10$^{11}$ L$_{\odot}$) than classical SMGs, are the host galaxies of $z\gtrsim$6 quasars (e.g.~\citealt{Walter09}, \citealt{Venemans12},\citeyear{Venemans18}, \citealt{Wang13}, \citealt{Decarli18}).

Quasars are among the most luminous sources in the universe; in recent years, the number of such objects known at $z\gtrsim$6 greatly increased, thanks to the advent of deep, large-area sky surveys (e.g.~\citealt{Fan06}, \citealt{Banados16}, \citeyear{Banados18}, \citealt{Mazzucchelli17b}, \citealt{Reed17}, \citeyear{Reed19}, \citealt{WangF18}, \citealt{Matsuoka18}).
Observations of the stellar light from their host galaxies has been very challenging, due to the much brighter, non-thermal radiation from the central engine (e.g.~\citealt{Mechtley12}, \citealt{Decarli12}).
On the other hand, emission from cool gas and dust in the observed (sub-)mm wavelength regime has been studied in several sources, providing a wealth of information on the composition, dynamics and conditions in the interstellar medium (ISM) of their hosts (e.g.~\citealt{Maiolino09}, \citealt{Willott15}, \citealt{Venemans16}, \citeyear{Venemans17}).
In particular, the singly ionized 158 $\mu$m carbon emission line, [\cii], is one of the main coolants of the ISM and is very bright (it can emit up to 1\% of the total far infrared emission in star--forming galaxies). It has been used extensively as a key diagnostic of galactic physics (see \citealt{Carilli13} and \citealt{DiazSantos17} for reviews, and \citealt{HerreraCamus18a}, \citeyear{HerreraCamus18b} for recent works).

Recently, \cite{Decarli18} and \cite{Venemans18} undertook a survey of [\cii] and underlying dust continuum emission in 27 quasar host galaxies at $z\gtrsim$6, with the Atacama Large Millimeter Array (ALMA), at a resolution of 1\arcsec, i.e.$\sim$5.5 pkpc at those redshifts.
Surprisingly, they serendipitously discovered [\cii]-- and far infrared--bright companion galaxies in the fields of four quasars, with projected separations of $\lesssim$60 kpc and line-of-sight velocity shifts of $\lesssim$450 km s$^{-1}$ \citep{Decarli17}.
In addition, \cite{Willott17} used ALMA observations at 0.\arcsec7 resolution (i.e.$\sim$4 pkpc at $z\sim$6.5) to find a very close companion galaxy to the quasar PSO J167.6415--13.4960 at $z\sim$6.5, at a projected distance of only 5 kpc and velocity separation of $\sim$300 km s$^{-1}$. 
Similar sources have also been observed in lower redshift systems (e.g.~at $z\sim5$; \citealt{Trakhtenbrot17}).
These findings, together with the discovery of a couple of galaxies adjacent to two quasars at $z\sim$4 and 6 \citep{McGreer14}, a Ly$\alpha$--emitting galaxy $\sim$12 kpc away from a $z\sim6.6$ quasar \citep{Farina17}, and a close quasar--galaxy pair at $z\sim$6 (Neeleman et al.~2019),
provide observational support to the theoretical paradigm that $z\sim$6 quasars reside in rich galactic environments (e.g.~\citealt{Volonteri06}, \citealt{Overzier09}, \citealt{Angulo12}). However, we note that other studies did not find overdensities of [\cii]/dust continuum--emitting galaxies (e.g.~\citealt{Venemans16}, \citealt{Champagne2018}), or of LAEs (e.g.~\citealt{Banados13}, \citealt{Mazzucchelli17a}, \citealt{Ota18}) around a sample of $z\gtrsim$6 quasars.
The observed [\cii]--bright companion galaxies have been considered as potential progenitors of $z\sim$4 red-and dead galaxies \citep{Decarli17}.
Previous optical/NIR observations have failed to detect rest-frame UV/optical emission from any of these companion galaxies, suggesting that they are heavily obscured and limiting the study of their overall physical properties \citep{Decarli17}.

In this work, we present new sensitive optical/NIR follow-up observations obtained from several ground- and space-based facilities, specifically designed to probe companion galaxies to four $6< z <6.6$ quasars.
In particular, we aim to observe the bulk of their stellar emission in the rest-frame optical wavelength range ($\sim$5000--7000 \AA), in order to assess their total stellar mass (M$_{*}$).
We also aim to measure their rest--frame UV radiation ($\sim$1200--1500 \AA), to probe the contribution from the young stellar population, and to determine how much of the star formation is unobscured.
We observed the fields around three quasars presented in \cite{Decarli17}: SDSS J0842+1218, PSO J231.6576--20.8335 and  CFHQS J2100$-$1715 (hereafter J0842, PJ231 and J2100, respectively), and around PSO J167.6415--13.4960 (hereafter PJ167; \citealt{Venemans15b}, \citealt{Willott17}).
In the following sections, we will refer to each of the respective companions as ``quasar\_short\_name''c.
We also obtained data for a mm--bright source, detected only in the dust continuum emission, close to the quasar VIK J2211$-$3206 (hereafter J2211; \citealt{Venemansinprep})\footnote{This quasar was also recently independently discovered by  \cite{Chehade18}, with the name of VST-ATLAS J332.8017-32.1036.}.
This galaxy is part of the sample of dust continuum--emitting sources discovered around several $z\sim$6 quasars by \cite{Champagne2018}, for which no redshift confirmation is available.
We present our follow--up data and discuss our constraints on the properties of this source in Appendix \ref{appJ2211}.

This paper is structured as follow:
In \S \ref{secObs} we present our observations and data reduction;
In \S \ref{secAnalysisSED} we compare the companion galaxies' photometry with the Spectral Energy Distributions (SEDs) of local galaxies, 
and in \S \ref{secAnalysisSFRUV} we estimate the (un-)obscured star formation rates from the rest frame (UV)optical emission.
In \S \ref{secAnalysisMSFR} we place the $M_{*}$ and SFR of the companions in the context of observations of SMGs and normal star--forming galaxies at comparable redshifts.
Finally, in \S \ref{secConc} we present our conclusions and outlook.

The magnitudes reported in this work are in the AB system. We use a $\rm \Lambda$CDM cosmology with H$\rm _{0}=$70 km s$^{-1}$ Mpc$^{-1}$, $\rm \Omega_{m}=$0.3, and $\rm \Omega_{\Lambda}=$ 0.7.
\section{Observations and Data Reduction} \label{secObs}
We collect available observations of the fields in our sample, either from the literature or obtained with dedicated follow-up campaigns. The coordinates, redshifts, spatial and velocity separations of the quasars and their companion galaxies are reported in Table \ref{tabSample}. Details on the optical/NIR observations used here, i.e.~dates, instruments/telescopes, exposure times and filters, are shown in Table \ref{tabObserv}.
\begin{deluxetable*}{lcccccccc}[h]
\tabletypesize{\small}
\tablecaption{Coordinates, redshifts, spatial projected distances and velocity shifts of the quasars and the adjacent galaxies studied in this work. These measurements are obtained from the narrow [\cii] emission line and underlying dust continuum observed by ALMA.
References are as: (1) \cite{Decarli17}, (2) \cite{Decarli18}, (3) \cite{Willott17} and (4) Neeleman et al.~2019.
\label{tabSample}}
\tablewidth{0pt}
\tablehead{ \colhead{name}& \colhead{R.A.~(J2000)} & \colhead{Decl.~(J2000)} & \colhead{$z$} & \colhead{$z_{err}$} & \colhead{$\Delta$r$_{\mathrm{projected}}$} & \colhead{$\Delta$v$_{\mathrm{line\, of\, sight}}$} & \colhead{References}\\
 & & & & & [kpc] & [km s$^{-1}$] &
}
\startdata
SDSS J0842+1218 &  08:42:29.43 & 12:18:50.4  & 6.0760 & 0.0006 & &  & (1)\\
SDSS J0842+1218c &  08:42:28.95 & 12:18:55.1  & 6.0656 & 0.0007 & 47.7 $\pm$ 0.8  &-443 &  (1)\\
PSO J167.6415--13.4960   &  11:10:33.98  & --13:29:45.6 & 6.5154 & 0.0003 &  &   & (4)\\
PSO J167.6415--13.4960c   &  11:10:34.03 & --13:29:46.3  & 6.5119 & 0.0003  & 5.0 & -140 & (4)\\
PSO J231.6576--20.8335   &  15:26:37.84 & --20:50:00.8 & 6.58651& 0.00017 &  &   & (1)\\
PSO J231.6576--20.8335c   &  15:26:37.87 & --20:50:02.3 & 6.5900 & 0.0008  & 8.4 $\pm$ 0.6 & +137 & (1) \\
CFHQS J2100$-$1715 & 21:00:54.70 & --17:15:21.9 & 6.0806 & 0.0011 &  & & (1)\\
CFHQS J2100$-$1715c & 21:00:55.45 & --17:15:21.7 & 6.0796 & 0.0008 & 60.7 $\pm$ 0.7 & -41  & (1)
\enddata
\end{deluxetable*}
\begin{deluxetable*}{lcccccc}[h]
\tabletypesize{\small}
\tablecaption{Information on optical/IR spectroscopic and imaging observations used in this work. 
Observations of the dust--continuum detected source close to the quasar VIK J2211$-$3206 are described in Appendix \ref{appJ2211}.
\label{tabObserv}}
\tablewidth{0pt}
\tablehead{ \colhead{name}& \colhead{Date/Program ID} & \colhead{Telescope/Instrument} & \colhead{Filters/$\lambda$ range} & \colhead{Exp.~Time}
}
\startdata
SDSS J0842+1218\footnote{Archival \textit{Spitzer}/IRAC [5.8],[8.0] and $HST$/WFC3 F105W data are taken from \citealt{Leipski14}.} & 2016--05--8/10  & LBT/MODS & 0.51--1.06 $\mu$m & 1320s\\ 
 & 2016--03--15 & Magellan/FIRE & 0.82--2.49 $\mu$m & 4176s \\ 
 & 2017--04--27 / 14876 & $HST$/WFC3 & F140W & 2612s\\
 & 2011--01--22 / 12184 & $HST$/WFC3 & F105W & 356s \\ 
 & 2017--02--09 / 13066 & $Spitzer$/IRAC  & 3.6, 4.5 $\mu$m & 7200s\\
 & 2007--11--24 / 40356 & $Spitzer$/IRAC  & 5.8, 8 $\mu$m & 1000s \\
PSO J167.6415--13.4960 & 2017--08--11 / 14876 & $HST$/WFC3 & F140W &  2612s\\
 & 2017--04--13 / 13066 & $Spitzer$/IRAC  & 3.6, 4.5 $\mu$m & 7200s \\
PSO J231.6576--20.8335 & 2017--07--02 / 099.A-0682 & VLT/MUSE & 0.465--0.93 $\mu$m & 10656s\\
 & 2016--03--15 & Magellan/FIRE & 0.82--2.49 $\mu$m & 4788s \\
 & 2017--04--01 / 14876 & $HST$/WFC3 & F140W & 2612s \\
 & 2016--11--25 / 13066 & $Spitzer$/IRAC  & 3.6, 4.5 $\mu$m & 7200s  \\
CFHQS J2100$-$1715 & 2016--08--25/26 / 297.A-5054 & VLT/MUSE & 0.465--0.93 $\mu$m & 7956s \\
 & 2016--09--18/19 / 334041 & LBT/LUCI & $J$ & 10440s\\
 & 2017--05--04 / 14876 & $HST$/WFC3 & F140W & 2612s \\
 & 2017--01--14 / 13066 & $Spitzer$/IRAC  & 3.6, 4.5 $\mu$m & 7200s \\
VIK J2211$-$3206 & 2017--04--28 / 14876 & $HST$/WFC3 & F140W & 2612s \\
 & 2017--01--29 / 13066 & $Spitzer$/IRAC  & 3.6,4.5 $\mu$m & 7200s  
\enddata
\end{deluxetable*}
\subsection{Optical/NIR Spectroscopy} \label{secObsSPEC}
We collected optical and NIR spectroscopic data for the quasars and their respective companions.\\
We observed the quasars PJ231 and J2100 with the Multi Unit Spectroscopic Explorer (MUSE; \citealt{Bacon10}) at the Very Large Telescope (VLT), imaging a total field of view of 1$\times$1 arcmin$^{2}$, with a spatial resolution of 0.2\arcsec/pixel and a spectral coverage between 4650-9300 \AA.
We observed the field of the quasar J2100 using Director's discretionary time (Program ID: 297.A-5054(A), PI: Decarli) during the nights of 25 and 26 August 2016. Sky conditions were good, with seeing varying from 0.8\arcsec~to 1.3\arcsec.
PJ231 was observed on July 2nd 2017 as a part of our program 099.A-0682A (PI: Farina) in almost photometric sky conditions and median seeing of 0.8\arcsec. We reduced the data using the MUSE Data Reduction Software (\citealt{Weilbacher12}, \citeyear{Weilbacher14}).
The final cubes were then post-processed as in \cite{Farina17}. In particular, the pipeline-produced variance cube was rescaled to match the observed variance of the background at each wavelength channel. This allowed us to compute more realistic errors that reflect possible correlations between neighboring voxels.
The spectrum of J2100c was extracted with a fixed aperture of 1\arcsec~radius centered at the position derived from our ALMA data. In PJ231, the quasar and companion are separated by only 1.5\arcsec, requiring careful removal of the quasar contribution. We created a Point Spread Function (PSF) model directly from the quasar by collapsing the spectral region $>$2000 km s$^{-1}$ redward of the Ly$\alpha$ line, at wavelength not contaminated by strong sky emission. At each wavelength channel, the PSF model was rescaled to match the flux of the quasar within 2 spaxels (0.4\arcsec) of the central emission and then subtracted. The spectrum of the companion was then extracted from this PSF--subtracted datacube with an aperture of 1\arcsec radius.
A more detailed analysis of this full dataset will be presented in Farina et al.~(in prep).

We also acquired spectra of the quasar J0842 with the Multi-Object Double Spectrograph (MODS; \citealt{Pogge10}) at the Large Binocular Telescope (LBT), in binocular mode on 8 and 10 May 2016.
The orientation of the slit covered both the quasar and the [\cii] companion galaxy. We used the 1.\arcsec2 slit and the GG495 filter. We collected two exposures of 1320s, for a total of 1hr28min on target. Data reduction was performed with standard Python and IRAF procedures. In particular, we corrected for bias and flat with the modsCCDRed package\footnote{http://www.astronomy.ohio-state.edu/MODS/Software/modsCCDRed/} and we wavelength-- and flux--calibrated the data using IRAF. The wavelength scale was calibrated using bright sky emission lines, delivering an accuracy of $\sim$0.2\AA~at $\lambda >$7000\AA. The standard star GD153 was observed to flux--calibrate the data.
We further scale the spectrum of the quasar J0842 to match the observed $z_{\mathrm{P1}}$ magnitude, 
as taken from the internal final release, PV3, of the Pan-STARRS1 Survey
($z_{\mathrm{P1}}=$19.92$\pm$0.03, \citealt{Magnier16}; see also \citealt{Jiang15} for further details on the discovery and the photometry of this quasar).
We applied this scaling to the spectrum of the companion, as extracted with a boxcar filter at the position obtained from the ALMA data.

\begin{figure}[!h]
\centering
\includegraphics[width=\columnwidth]{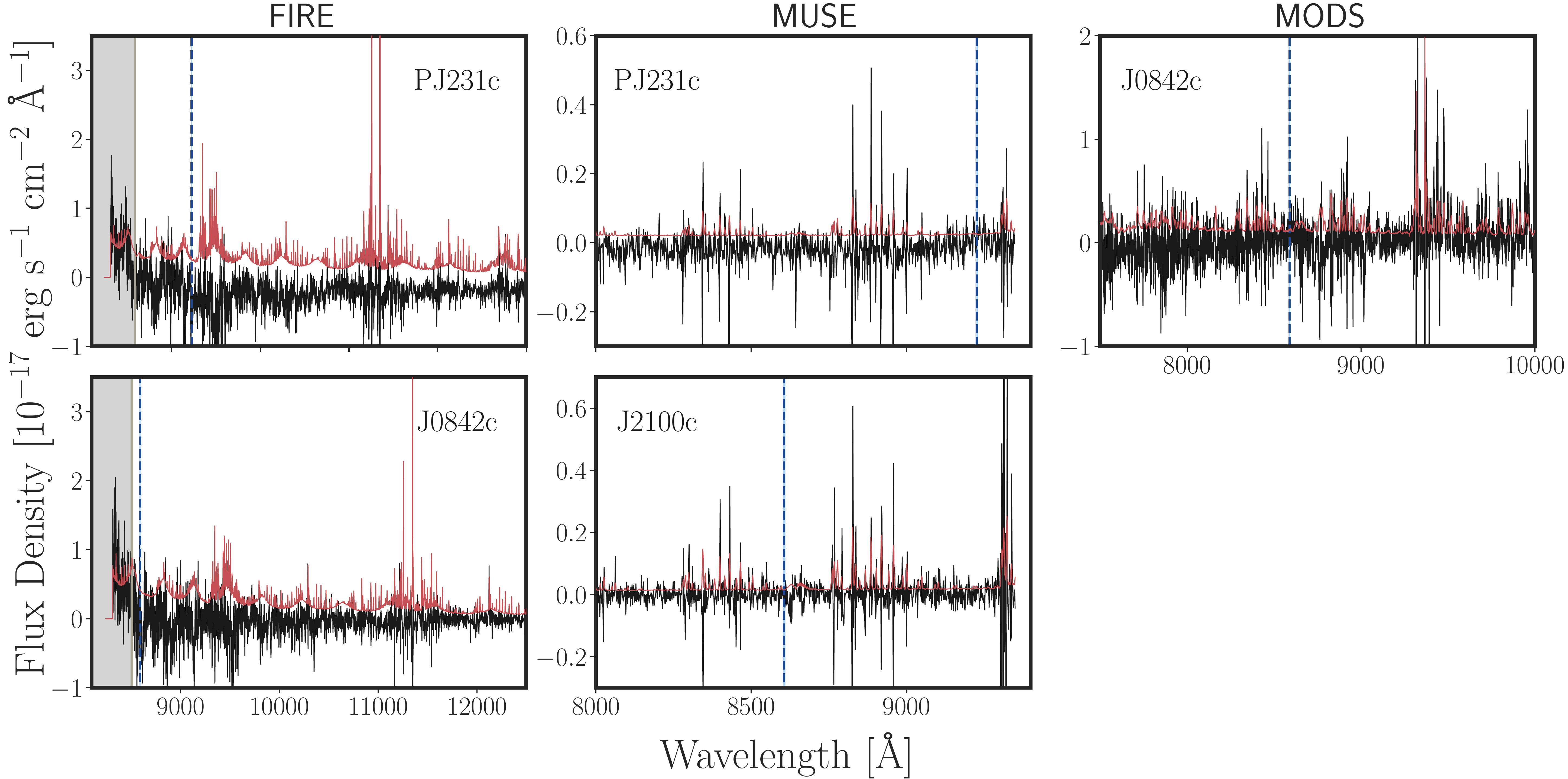}
\caption{
Spectra at the locations of the companions to PJ231, J2100 and J0842, acquired with the FIRE and MODS spectrographs, and extracted from the MUSE datacubes.
We highlight the spectral regions where the flux calibration is less reliable with \textit{grey shaded areas}.
\textit{Dashed blue lines} highlight the expected positions of the respective Ly$\alpha$ emission lines, established from the observations of the narrow [\cii] emission with ALMA. The surrounding regions of $\pm$100 km/s (rest--frame), used to estimate limits on the Ly$\alpha$ emission line in the companion galaxies, are also shown with \textit{light blue shaded areas}. 
}
\label{figSPECFIRE}
\end{figure}

We observed the companions of PJ231 and J0842, and the quasar PJ231, with the Folded-port InfraRed Echelette (FIRE; \citealt{Simcoe08}) at the Magellan Baade Telescope. We observed PJ231 and its companion simultaneously, while we performed a blind offset from the quasar to J0842c.
The data were reduced following standard techniques, including bias subtraction, flat field and sky subtraction.
The wavelength calibration was obtained using sky emission lines as reference (see also \citealt{Banados14}).
We used the standard stars HIP43018 and HIP70419 to flux calibrate and correct for telluric contamination in the spectra of J0842c and PJ231c, respectively; we implemented the absolute flux calibration considering the $J$ magnitude of PJ231 ($J$ AB=19.66$\pm$0.05; \citealt{Mazzucchelli17b}).\\
We show all the spectra extracted at the companion positions in Figure \ref{figSPECFIRE}.
No clear emission from any of the companions spectra is detected.
In all cases, we estimated the 3$\sigma$ limits on the Ly$\alpha$ emission line as:
\begin{equation}
    F_{\mathrm{Ly\alpha},3\sigma} = 3 \times \sqrt{\sum_{i=1}^{N} err_{i}^{2}} \times \sum_{i=1}^{N} \Delta \lambda_{i} 
\end{equation}
where \textit{err} is the error vector, N is the number of pixels in within a velocity window of (rest--frame) 200 km/s (i.e.~the typical line width measured in LAEs, e.g.~\citealt{Ouchi08}) around the supposed location of the Ly$\alpha$ emission line (as obtained from the ALMA [\cii] observations).
Finally, $\Delta \lambda = \lambda_{i+1} - \lambda_{i}$ in the considered wavelength interval.
Moreover, we calculated the limits on the underlying continuum emission as:
\begin{equation}
    F_{\mathrm{cont},3\sigma} = 3 \times \sqrt{\sum_{i=1}^{N} err_{i}^{2}}
\end{equation}
where we consider here the spectral coverage at hands, excluding noisy regions at the edges.
All the estimated values are reported in Table \ref{tabSPECdata}.

We note that the emission from the Ly$\alpha$ line in $z\sim$6--7 LAEs can be redshifted by $\sim$100-200 km/s with respect to the [\cii] line, and/or it can be originated from slightly different spatial locations (e.g.~\citealt{Pentericci16}).
Here, the limits we measure by shifting the center of the Ly$\alpha$ emission by $\pm$150 km/s are consistent with the fiducial values reported in Table \ref{tabSPECdata}, i.e.~we do not significantly detect a blue/redshifted line.
\begin{figure}[!h]
\centering
\includegraphics[width=\columnwidth]{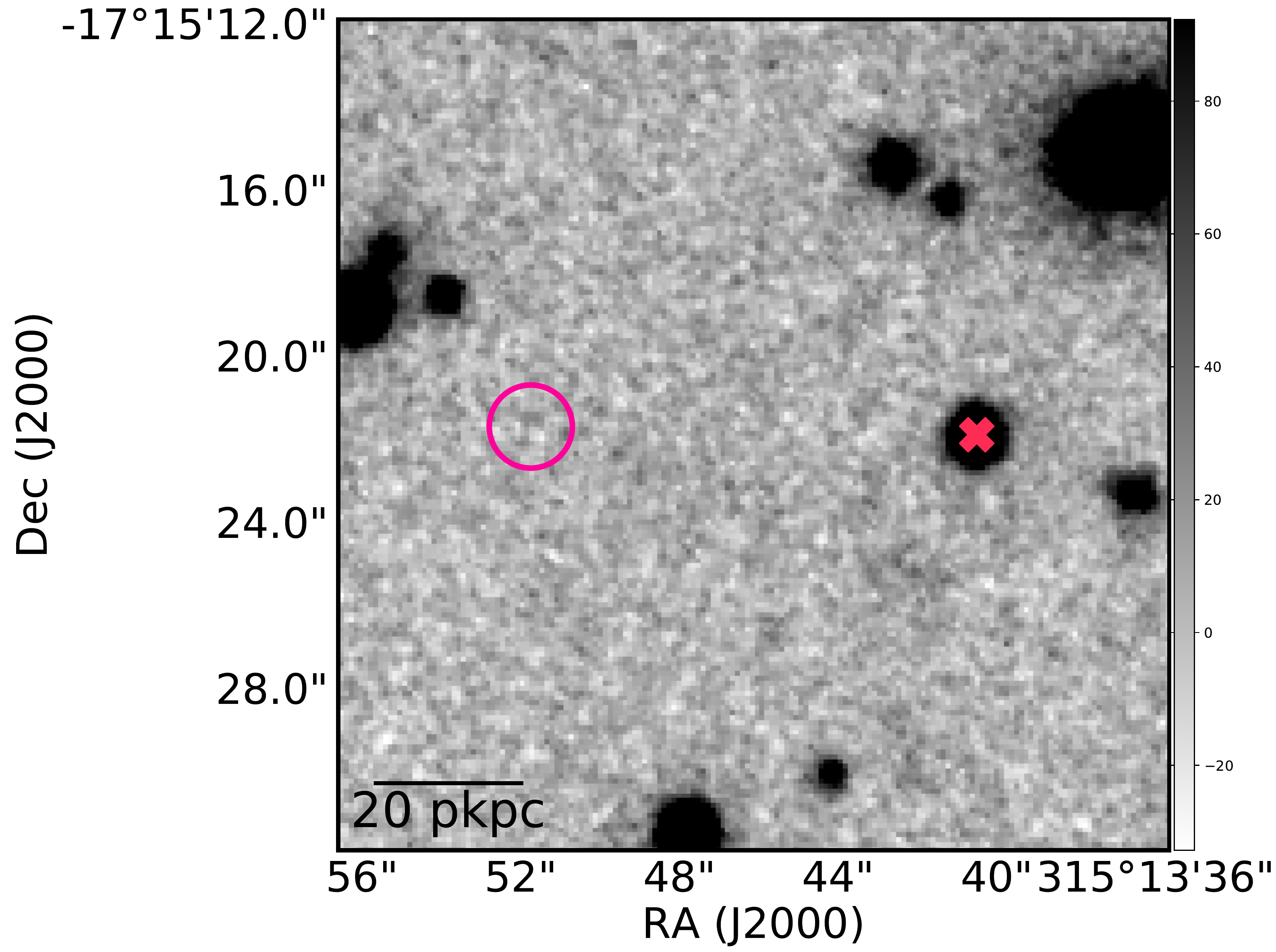}
\caption{Postage stamp (20\arcsec$\times$20\arcsec) of the field around the quasar J2100, imaged in the \textit{J} filter with the LUCI1 and LUCI2 cameras at the LBT (see Section \ref{secObsLUCI} and Table \ref{tabObserv}). We place a limit of $J>$26.24 (at 3$\sigma$) on the emission from the companion galaxy (\textit{magenta circle}).
}
\label{figLBTLUCI}
\end{figure}
\begin{deluxetable*}{lccc}
\tabletypesize{\small}
\tablecaption{Measurements of the strength of Ly$\alpha$ emission line and of the underlying rest--frame UV continuum from the spectroscopic observations of the companion galaxies to J0842, PJ231 and J2100, obtained with VLT/MUSE, Magellan/FIRE and LBT/MODS. Limits are at 3$\sigma$ significance, and obtained as described in Section \ref{secObsSPEC}. \label{tabSPECdata}}
\tablewidth{0pt}
\tablehead{
\colhead{name} & \colhead{\underline{VLT/MUSE}} & \colhead{\underline{LBT/MODS}} & \colhead{\underline{Magellan/FIRE}}
\vspace{1mm}\\ 
 & $F_{\mathrm{Ly\alpha}} \quad F_{\mathrm{cont}}$ & $F_{\mathrm{Ly\alpha}} \quad F_{\mathrm{cont}}$ & $F_{\mathrm{Ly\alpha}} \quad F_{\mathrm{cont}}$ 
\vspace{0.6mm}\\
& [erg/s/cm$^{2}$] [erg/s/cm$^{2}$/\AA] & [erg/s/cm$^{2}$] [erg/s/cm$^{2}$/\AA] & [erg/s/cm$^{2}$] [erg/s/cm$^{2}$/\AA]
}
\startdata
SDSS J0842+1218 & -- & $<$4.5e-17 $\quad$ $<$3.0e-16 & $<$2.7e-16 $\quad$ $<$8.5e-16 \\ 
PSO J231.6576--20.8335 & $<$2.1e-16 $\quad$ $<$1.4e-15 & -- & $<$1.8e-16 $\quad$ $<$8.5e-16 \\
CFHQS J2100$-$1715 & $<$8.3e-17 $\quad$ $<$3.5e-15 & -- & --
\enddata
\end{deluxetable*}
\subsection{IR Photometry}
We list here the observations and data reduction of the imaging follow--up data, obtained with ground-- and space--based instruments.
\subsubsection{LUCI @ LBT} \label{secObsLUCI}
We imaged the field of J2100 in the $J$ band ($\lambda_{c}=$1.247 $\mu$m and $\Delta \lambda=$0.305 $\mu$m) with the Utility Camera in the Infrared (LUCI1 and LUCI2; \citealt{Seifert03}) at the LBT, in binocular mode.
We reduced the data following standard techniques, i.e.~we subtracted the master dark, divided by the master flat field, and median--combined the frames after subtracting the contribution from the background and after aligning them using field stars. The final astrometric solution used the Gaia Data Release 1 catalog\footnote{https://www.cosmos.esa.int/web/gaia/dr1} (DR1; \citealt{GaiaDR16a}, \citeyear{GaiaDR16b}) as reference.
We flux-calibrated the image with respect to the 2MASS Point Source Catalog. The seeing of the reduced image is 0.\arcsec98.
We calculated the depth of the image by distributing circular regions with radius equal to half of the seeing over the frame, in areas with no sources. The 1$\sigma$ error of our image is the standard deviation of the Gaussian distribution of the fluxes calculated in these apertures. 
We do not detect, at S/N$>$3, any emission at the location of the companion, after performing forced photometry in a circular aperture whose diameter is corresponding to the seeing (see Figure \ref{figLBTLUCI}).
The 3$\sigma$ limit magnitude that we will use in the following analysis is $J$=26.24 ($F_{J}$=0.116 $\mu$Jy; see Table \ref{tabCOMPPhot}).
\subsubsection{WFC3 @ \textit{HST}} \label{secObsHST}
\begin{figure*}
\centering
\includegraphics[width=\textwidth]{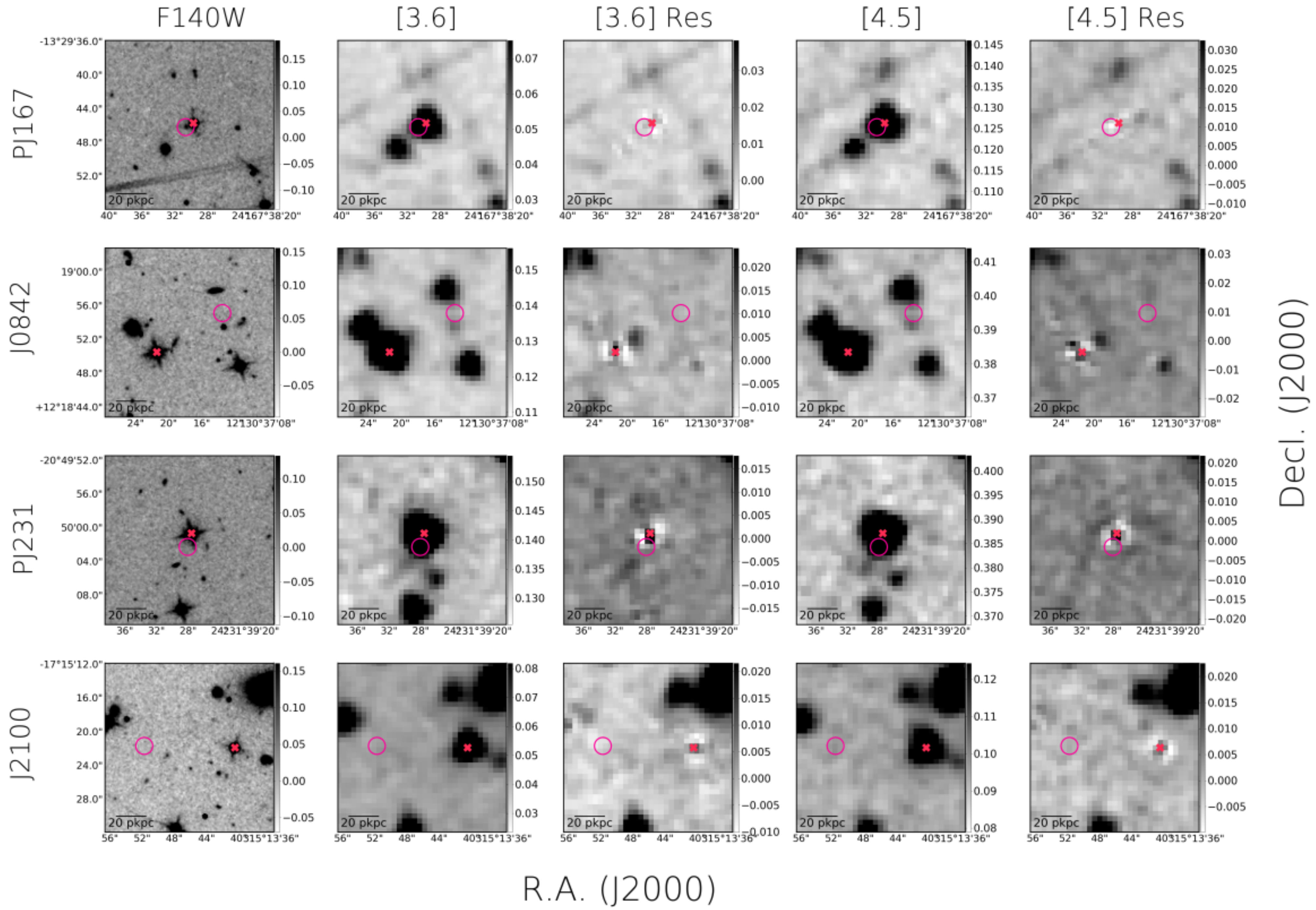}
\caption{Postage stamps (20\arcsec$\times$20\arcsec) of the four fields (quasar+companion) considered in this study.
We also report the residual IRAC images after removing the emission from the quasar and nearby foreground sources (see Section \ref{secObsSPITZER}).
The positions of the companions and of the quasars are highlighted with \textit{magenta circles} (of 1\arcsec radius) and \textit{red crosses}, respectively.
}
\label{figPSall}
\end{figure*}

\begin{figure*}
\centering
\includegraphics[width=0.7\textwidth]{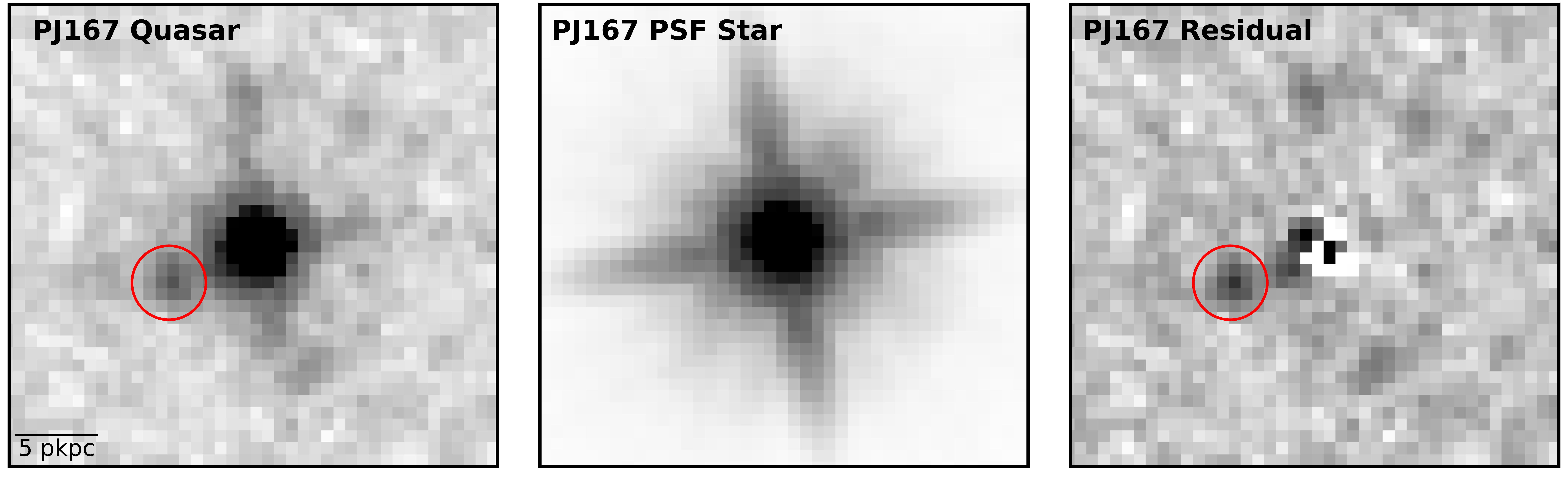}
\caption{\textit{HST}/WFC3 image in the F140W filter of the quasar PJ167.
\textit{Left panel}: native postage stamp (5\arcsec$\times$5\arcsec).
\textit{Central panel}: empirical PSF model obtained from a bright star in the field (see Section \ref{secObsHST}).
\textit{Right panel}: residual image after the subtraction of the empirical PSF from the data.
The companion galaxy observed in the ALMA image is detected and well resolved in the latter frame (\textit{white circle}, of radius 0.\arcsec4). Additional residual flux, located between the center of the bright quasar and the adjacent galaxy is also tentatively detected.
}
\label{figPSFPJ167}
\end{figure*}

We obtained new observations of all the targets studied here with the Wide Field Camera 3 (WFC3), on board \textit{HST}, using the F140W filter ($\lambda_{c}=$1.3923 $\mu$m and $\Delta \lambda=$0.384 $\mu$m;
Program ID:14876, PI: E.~Ba\~nados).
For the quasar J0842, previous WFC3 observations in the F105W filter ($\lambda_{c}=$1.0552 nm and $\Delta \lambda=$0.265 nm) were also retrieved from the Hubble Legacy Archive\footnote{https://hla.stsci.edu/} (Program ID:12184,PI: X.~Fan).
We refer to Table \ref{tabObserv} for further details on this dataset.
We analyzed both the archival and new observations in a consistent way. 
We considered the reduced data produced by the \textit{HST} pipeline, and we took the zero-point photometry from the WFC3 Handbook\footnote{http://www.stsci.edu/hst/wfc3/analysis/ir\_phot\_zpt}.
We re-calibrated the astrometry using the Gaia DR1 catalog (see also Section \ref{secObsLUCI}).
We calculated the depth of the images in an analogous way as in Section \ref{secObsLUCI}, considering here circular areas of 0.4\arcsec radius (containing the 84\% of the flux from a point source\footnote{http://www.stsci.edu/hst/wfc3/analysis/ir\_ee}). 
We performed aperture photometry using this aperture radius at the positions of the companions.
The companion sources of J0842, J2100 and PJ231 were not detected in the F140W filter, and
J0842c was also not detected in the F105W image. We report all the 3$\sigma$ limit fluxes in Table \ref{tabCOMPPhot}.
We show the observations of all the fields studied in this work in the F140W filter in Figure \ref{figPSall}, and the F105W image of J0842 in Figure \ref{figJ0842DB}.
In the case of PJ167, the companion is located at a projected distance of only 0.9\arcsec, and it is blended with the quasar emission.
In order to recover meaningful constraints on the brightness of PJ167c, it is necessary to subtract the quasar contribution by modeling the image PSF. 
We used the bright star 2MASS J11103221--1330007, in the proximity of PJ167,
in order to create an empirical PSF model from the same image. This source is located at a distance of only 30\arcsec~from the quasar, limiting the errors due to the changes in the PSF over the field.
Its $J$ and $H$ magnitudes from the 2MASS Point Source Catalog are 15.249 and 15.105, respectively.
The corresponding $J-H$ color of 0.144 is therefore close to that of the quasar ($J-H$=0.216).
We shifted, scaled and subtracted the PSF model from the quasar emission using the software \texttt{GALFIT} (version 3.0.5; \citealt{Peng02}, \citeyear{Peng10}). In Figure \ref{figPSFPJ167} we show the native \textit{HST} image, the PSF star model, and the residual frame, in which the bright quasar emission has been subtracted.
The companion galaxy is well isolated, and its F140W PSF magnitude, measured with \texttt{GALFIT}, is equal to 25.48 $\pm$ 0.17 ($F_{\mathrm{F140W}}$=0.23$^{+0.04}_{-0.03}$ $\mu$Jy).
Diffuse emission extending from the companion to the quasar is also tentatively recovered. Additional, high resolution imaging and spectroscopy are needed to securely confirm and characterize such emission.
\vspace{0.4pt}

\subsubsection{IRAC @ \textit{Spitzer}} \label{secObsSPITZER}
The fields of all the objects in our sample were observed in the [3.6]
($\lambda_{c}=$3.550 $\mu$m and $\Delta \lambda=$0.750 $\mu$m)
and [4.5]
($\lambda_{c}=$4.493 $\mu$m and $\Delta \lambda=$1.015 $\mu$m)
filters with the InfraRed Array Camera (IRAC; \citealt{Fazio04}; Program ID:13066, PI: C.~Mazzucchelli; see Table \ref{tabObserv}).
We also use archival data of J0842 (Program ID:40356, PI: X.~Fan), covering the IRAC filters [5.8] ($\lambda_{c}=$5.731 $\mu$m and $\Delta \lambda=$1.425 $\mu$m)
and [8.0] ($\lambda_{c}=$7.872 $\mu$m and $\Delta \lambda=$2.905 $\mu$m; see Table \ref{tabObserv}).

We adopt the reduced data from the \textit{Spitzer} pipeline, and the photometric calibration (i.e.~photometric zero point and aperture correction values) specified in the IRAC Instrument Notebook\footnote{http://irsa.ipac.caltech.edu/data/SPITZER/docs/
irac/iracinstrumenthandbook/IRAC\_Instrument\_Handbook.pdf}.
As in the case of the \textit{HST}/WFC3 observations, we refine the astrometric solution using the recent Gaia DR1 catalog.
Given the limited spatial sampling of the IRAC camera (0.6 arcsec/pixel) and the depth of our [3.6] and [4.5] images, the companion galaxies studied here are blended either with the much brighter quasar, or with foreground sources (see Figure \ref{figPSall}).
Hence, one needs to properly model and remove these sources.
In order to model the PSF, which is undersampled in the IRAC data, we re-sample the native images over a grid of 0.12 arcsec/pixel resolution, using the IRAF task \textit{magnify}.
In each magnified image, we select a collection of stars identified as such in the \textit{HST}/WFC3 data within 1\arcmin $\times$1\arcmin~of the quasar.
We obtain the final PSF model for each image by shifting, aligning, scaling and combining the images of the selected stars. The number of stars used in each field ranges between 4 and 9.
We use \texttt{GALFIT} to sample the PSF image to the original resolution (0.6 arcsec/pixel), and to model and subtract the emission from the quasar and any foreground objects.
In Figure \ref{figPSall}, we show the postage stamps of the IRAC [3.6] and [4.5] images, and the corresponding images of the residuals.
No clear emission from the companion galaxies is detected in any of the residual images.
We quantify the limits on the photometry of the companions as follows. 
For each image, we run \texttt{GALFIT} subtracting a source at the exact position of the companion modeled as a PSF and scaled to a fixed magnitude, which we vary between 21 and 25, in steps of 0.01 mag.
When adopting magnitudes smaller (i.e.~brighter fluxes) than the limit flux to which our image is sensitive, the subtraction will leave a negative residual.
We perform aperture photometry in the residual image at the companion position in an aperture of 2.4\arcsec~radius, and we compare the measured flux with the image 3$\sigma$ flux limit. The adopted flux limit was measured on the same area used for the forced photometry, and by evaluating the background rms in an annulus of radius 14\arcsec~and width of 10\arcsec~centered on the companion.
We assume that the 3$\sigma$ limit magnitude is the value at which the measured absolute flux in the residual image is equal to the 3$\sigma$ flux limit.
We report these values in Table \ref{tabCOMPPhot}.

\begin{figure*}
\centering
\includegraphics[width=0.9\textwidth]{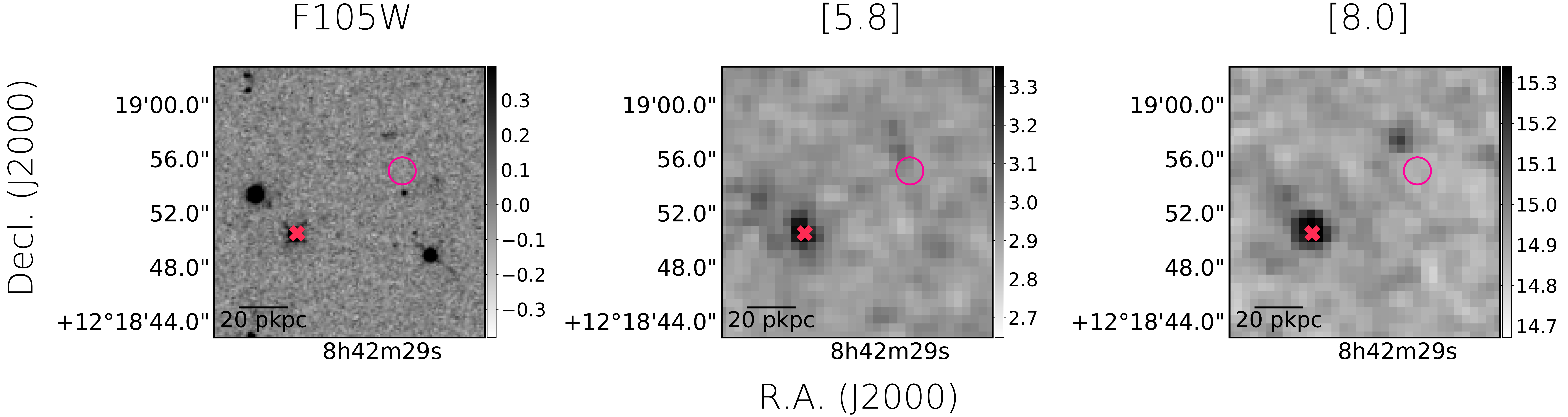}
\caption{
Archival observations (20\arcsec$\times$20\arcsec) of the field around the quasar J0842. The \textit{left} panel shows the image obtained from the \textit{HST}/WFC3 instrument in the F105W filter, while the other two panels show observations acquired with the \textit{Spitzer}/IRAC camera, in the [5.8] and [8.0] channels (see Section \ref{secObsSPITZER} and Table \ref{tabObserv} for references).
The quasar is identified with a \textit{red cross}, while the companion position is highlighted with a \textit{magenta circle}.
These observations were acquired with the aim of studying the bright quasar emission, therefore the flux limits at the companion position are shallower than the newly obtained images (see Table \ref{tabCOMPPhot}).
}
\label{figJ0842DB}
\end{figure*}

Finally, we analyze the archival J0842 \textit{Spitzer}/IRAC observations (see Figure \ref{figJ0842DB}), which are much shallower (see Table \ref{tabObserv}), since they were designed only to detect the bright quasar.
No foreground objects overlap the companion location, and we therefore perform aperture photometry on the native images, using the same aperture as in the observations in the [3.6] and [4.5] channels.
We do not detect any source at S/N$>$3. We report the corresponding flux limits in Table \ref{tabCOMPPhot}.
\\[1mm]

\begin{deluxetable*}{lcccccccccc}
\tabletypesize{\small}
\tablecaption{Photometric measurement of the companion galaxies to $z\sim6$ quasars studied in this work (see Section \ref{secObs}).
The limits provided are at 3$\sigma$ significance
\label{tabCOMPPhot}}
\tablewidth{0pt}
\tablehead{
\colhead{name} & \colhead{$F_{J}$} & \colhead{$F_{\mathrm{F105W}}$} & \colhead{$F_{\mathrm{F140W}}$} & \colhead{$F_{\mathrm{3.6  \mu m}}$} & \colhead{$F_{\mathrm{4.5  \mu m}}$} & \colhead{$F_{\mathrm{5.8  \mu m}}$} & \colhead{$F_{\mathrm{8.0  \mu m}}$} &
\colhead{$F_{\mathrm{1.2 mm}}$}
\\
 & [$\mu$Jy] & [$\mu$Jy] & [$\mu$Jy] & [$\mu$Jy] &  [$\mu$Jy] & [$\mu$Jy] &  [$\mu$Jy] & [mJy]
}
\startdata
SDSS J0842+1218c & -- & $<$0.154  &$<$0.061 & $<$0.78 & $<$1.06 & $<$9.54 & $<$12.6 & 0.36 $\pm$ 0.12 \\ 
PSO J167.6415--13.4960c & -- & -- & 0.23$^{0.04} _{0.03}$ & $<$0.78 & $<$1.28 & -- & -- & 0.16 $\pm$ 0.03\footnote{This flux measurement comes from recent 0.35\arcsec, i.e.~$\sim 2$ pkpc at $z \sim 6.5$, ALMA observations (Neeleman et al.~2019).} \\
PSO J231.6576--20.8335c & -- & -- & $<$0.053 & $<$0.64 & $<$2.79 & -- & -- & 1.73 $\pm$ 0.16\\
CFHQS J2100$-$1715c & $<$0.116 & -- & $<$0.083 & $<$0.53  & $<$1.07 & -- & -- & 2.05 $\pm$0.27 
\enddata
\end{deluxetable*}
\section{Analysis} \label{secAnalysis}
In this section, we characterize the SEDs of four companions to $z\sim6$ quasars, by comparing them with a few examples of local galaxies and by modeling their emission with a SED fitting code.
We estimate (or set upper limits to) their un--obscured/obscured star formation activity observed in the rest--frame UV/IR range. 
Finally, we place our measurements in the context of observations of star--forming galaxies and starbursts at similar redshift.

\subsection{\textit{Spectral Energy Distribution}} \label{secAnalysisSED}
We first compare the SEDs of our companions with those of prototypical galaxies in the local universe. 
We consider the SEDs of normal star forming spiral galaxies (M51 and NGC6946), starbursts (M82) and ultraluminous infrared galaxies (ULIRGs; Arp 220), from \cite{Silva98}.
M51 is a nearby ($D$=9.6 Mpc) spiral (Sbc) interacting galaxy, which has been studied in detail over a wide range of wavelength and physical scales (e.g.~\citealt{Leroy17}).
NGC6946, found at a distance of 6.72 Mpc, is an intermediate (Scd) spiral galaxy \citep{Degioia84}. Its size is approximately a third of that of our Galaxy and it hosts roughly half of the stellar mass (e.g.~\citealt{Engargiola91}).
M82 is a prototypical edge--on starburst (with a galaxy-wide SFR$\sim$ 10--30 M$_{\odot}$ yr$^{-1}$; \citealt{Forster2003}), whose intense activity has been most probably triggered by a past interaction with the neighboring galaxy M81 (e.g.~\citealt{Yun94}).
Arp 220 is one of the closest (77 Mpc) and best studied ULIRGs, with a total infrared luminosity of $L_{IR}=$1.91$\times$10$^{12}$ M$_{\odot}$ \citep{Armus09}.
It is thought to be the result of a merger which happened $\sim$3-5 Myr ago (e.g.~\citealt{JosephWright85}, \citealt{BaanHaschick95}, \citealt{Scoville98}, \citealt{DownesEckart07}),
and has extreme conditions in its nucleus (e.g.~with an attenuation of A$_{V}=2\times10^{5}$mag; \citealt{Scoville17})

\begin{figure*}[!h]
\centering
\includegraphics[width=\textwidth]{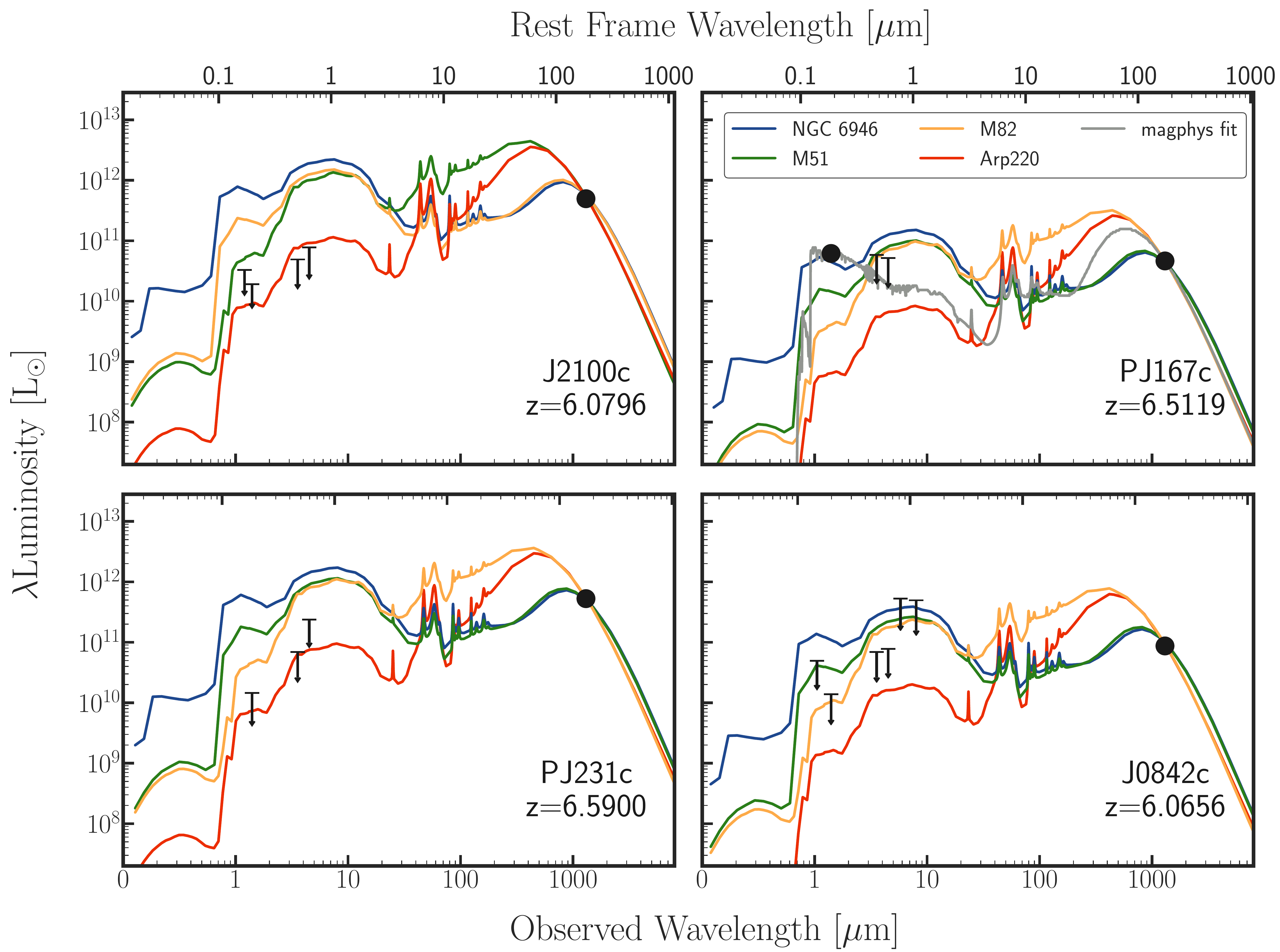}
\caption{Spectral Energy Distribution of four companion galaxies adjacent to $z\sim$6 quasars.
The measurements from our photometric observations (Table \ref{tabCOMPPhot}) are reported with \textit{down-pointing arrows} (limits at 3$\sigma$ significance) and \textit{filled black points}.
As comparison, we show representative SEDs of various local star forming galaxies (NGC 6946, \textit{blue}; M51, \textit{green}) and starbursts/ULIRG (M82, \textit{orange}; Arp 220, \textit{red line}; \citealt{Silva98}), normalized to the ALMA 1.2 mm measurement.
The best fit template (\textit{grey line}) of the SED of PJ167c, obtained with the code \texttt{MAPGPHYS}--highz \citep{DaCunha15}, is also reported.
The SEDs of J2100c, J0842c and PJ231c are consistent with being Arp 220 like-galaxies, i.e.~intensely forming stars and highly dust obscured, at $z\sim$6. The \textit{HST}/WFC3 measurement of the rest--frame UV emission of PJ167c suggests that this source is more similar to a ``regular'' starforming galaxy (e.g.~NGC6964), with a lower stellar mass.
}
\label{figSEDobs}
\end{figure*}

Here, we shift the observed SEDs of these local galaxies to the redshifts of the companions, and we scale them to match the 1.2mm flux retrieved in the ALMA observations.
We plot the SEDs, together with the photometry of the companions presented here, in Figure \ref{figSEDobs}.
The limits on rest--frame UV/optical brightness of PJ231c, J2100c and J0842c rule out all the galaxy templates considered here, with the exception of Arp 220.
These companions have infrared luminosities in the range of (ultra-) luminous galaxies (e.g.~0.9--5$\times$10$^{12}$ L$_{\odot}$; \citealt{Decarli17}).
On the other hand, the rest-frame UV emission of PJ167c is detected in our \textit{HST}/WFC3 observations (see Section \ref{secObs}).
Its UV-to-submm ratio is comparable to that of the star--forming galaxy NGC6946, while the limits from our \textit{Spitzer}/IRAC data suggest that it has a lower stellar content.

We compute the star formation rates for PJ231c, J2100c and J0842c assuming that their SEDs are equivalent to that of Arp 220, shifted in redshift and scaled as in Figure \ref{figSEDobs}.
We derive their star formation rates from the dust emission in the rest-frame infrared region (SFR$_{\rm IR}$), assuming that the non-obscured SFR as negligible (see Section \ref{secAnalysisSFRUV}).
We calculate the total IR luminosity by integrating the emission from 3 $\mu$m to 1100 $\mu$m, and we measure the SFR as: SFR$=1.49\times 10^{-10}~L_{\mathrm{IR}}$ \citep{Kennicutt12}.
The obtained values range between 120 and 700 M$_{\odot}$ yr$^{-1}$. 
We note that, if we had instead assumed a modified black body model, $f_{\nu} \propto B_{\nu} (T_{d} \nu^{\beta})$, and adopted typical parameters for high--redshift galaxies ($T_{d}=47$ K and $\beta=1.6$; e.g.~\citealt{Beelen06}, \citealt{Venemans16}), we would have derived comparable star formation rate values ($\sim$140--800 M$_{\odot}$ yr$^{-1}$; see also \citealt{Decarli17}).

We can obtain conservative upper limits on the companion stellar masses by subtracting their gas mass ($M_{\rm gas}$) from their dynamical mass ($M_{\rm dyn}$).
The latter can be obtained from the widths of the [\cii]~emission line observed with ALMA (see \citealt{Decarli17}).
We note, however, that these estimates are based on a number of important assumptions on the companions geometry and dynamics (i.e.~they are virialized systems), and that they are obtained from data with a relatively modest resolution of $\sim$1\arcsec.
We list all dynamical masses in Table \ref{tabSFRMstar}.
On the other hand, we can estimate $M_{\rm gas}$ from the dust content ($M_{\rm dust}$).
We take these values from \cite{Decarli17}: $M_{\rm dust}$ are measured following the prescription by \cite{Downes92}, while $M_{\rm gas}$ are obtained assuming a typical gas--to--dust ratio of $\sim$100 (e.g.~\citealt{Berta16}).
We obtain upper limits on the stellar content ranging between $\sim16$ and $21 \times 10^{10}$ M$_{\odot}$.
In the following analysis, we utilize the latter values as upper limits on the stellar masses of the companions (see Table \ref{tabSFRMstar} and Figures \ref{figMstarFobs} and \ref{figSFRMstarObs}).

Alternatively, we can compare our photometric measurements with synthetic galaxy templates.
We use the SED fitting code \texttt{MAGPHYS} \citep{DaCunha08}, which uses of an energy balance argument to combine simultaneously the radiation from the stellar component, the dust attenuation, and the re-emission in the rest-frame IR wavelength range.
We consider here the \texttt{MAPGPHYS}--highz extension \citep{DaCunha15}, which was specifically designed to characterize a sample of SMGs at $3 < z < 6$ (see also Section \ref{secAnalysisMSFR}). 
In particular, this version included younger galaxy templates, with higher dust extinction, and a wider choice of star formation histories. 
Nevertheless, fitting the photometry of the companion galaxies presented here with any code is hard, due to the few (and most of the time only one) broad--band detections for each source.
This is reflected in strong parameter degeneracies in the fit. Another issue is represented by the potentially inappropriate coverage of the parameter space considered in the fitting machine, which might not be modeling the properties of the peculiar galaxies considered here.
Therefore we choose to fit only the companion of PJ167, whose emission is retrieved in more than one broad band.
In Figure \ref{figSEDobs}, we show the best fit template from \texttt{MAGPHYS}--highz for this galaxy.
We take the 50th and 16th/84th percentiles of the marginalized probability distributions as the best fit values and uncertainties of its SFR and stellar mass.
The SED of PJ167c is consistent with that of a star forming galaxy, SFR$=$53$^{+27}_{-19}$ M$_{\odot}$ yr$^{-1}$, with a stellar mass of $M_{*}= 0.84^{+0.64}_{-0.40} \times 10^{10}$ M$_{\odot}$, a moderate dust extinction ($A_{V}=0.66^{+0.35}_{-0.25}$ mag) and a dust content of $M_{\rm d}= 4.7^{+3.7}_{-1.7} \times 10^{7}$ M$_{\odot}$.

Finally, we note that, given the close spatial/velocity separation of the companions and the quasar hosts, they are very likely found in physical connection. In particular, PJ167c is located at only 5 pkpc/140 km s$^{-1}$ away from PJ167, and emission linking these systems is observed both in the dust continuum and the [\cii] line (with a smooth velocity gradient; \citealt{Decarli17}, Neeleman et al.~2019) and, tentatively, in the rest--frame UV (see Fig.~\ref{figPSFPJ167}). This evidence, together with a measured high velocity dispersion of the cool gas ($\sim$150 km s$^{-1}$; Neeleman at al.~2019) may suggest that these galaxies have already entered an advanced
merging stage.
\begin{deluxetable*}{lcccccc}[h]
\tabletypesize{\small}
\tablecaption{
Physical properties of the companion galaxies to $z\sim$6 quasars studied in this work.
We report the unobscured (rest--frame UV) SFRs calculated from our $HST$/WFC3 observations (Section \ref{secAnalysisSFRUV}), and the obscured (rest--frame IR) contribution from our ALMA data (Section \ref{secAnalysisSED} and \ref{secAnalysisSFRUV}).
Finally, the dynamical mass estimates and upper limits on the stellar masses are also listed. 
In case of PJ167c, the reported stellar mass is that derived from \texttt{MAGPHYS}--highz (see Section \ref{secAnalysisSED}).
We note that the SFR$\rm _{[\cii]}$ values have an additional uncertainty of 0.5 dex due to the scatter around the relationship from \cite{DeLooze14}.
\label{tabSFRMstar}
}
\tablewidth{0pt}
\tablehead{ \colhead{name} & \colhead{SFR$\rm _{UV}$} & \colhead{SFR$\rm _{IR}$} & \colhead{SFR$\rm _{[\scii]}$} & \colhead{$f_{obscured}=$} & \colhead{M$\rm _{dyn}$} & \colhead{M$\rm _{*}$} \\
   & [M$_{\odot}$ yr$^{-1}$] & [M$_{\odot}$ yr$^{-1}$] & [M$_{\odot}$ yr$^{-1}$] & SFR$\rm _{IR}$/SFR$\rm _{UV+IR}$ & [$\times 10^{10}$ M$_{\odot}$]  & [$\times 10^{10}$ M$_{\odot}$]
}
\startdata
  SDSS J0842+1218c & $<$2 & 124 $\pm$ 54 & 260 $\pm$ 40 & $>$0.98 &  12 $\pm$ 5 & $<$11\\
  PSO J167.6415--13.4960c & 11 $\pm$ 3 & 32 $\pm$ 4 & 182 $\pm$ 16 & 0.74 $\pm$ 0.20 & -- & 0.84$^{+0.64}_{-0.40}$\\
  PSO J231.6576--20.8335c & $<$3  & 709 $\pm$ 157 & 730 $\pm$ 100 & $>$0.99 & 22 $\pm$ 8 & $<$16.8 \\
  CFHQS J2100$-$1715c & $<$3  & 573 $\pm$ 73 & 360 $\pm$ 70 & $>$0.99  & 27 $\pm$ 13 & $<$21.5
\enddata
\end{deluxetable*}
\subsection{SFR$_{UV}$ vs SFR$_{IR}$} \label{secAnalysisSFRUV}
The rest--frame UV emission of galaxies directly traces young stars, i.e.~10--200 Myr old. It is thus an excellent probe of recent star formation, but it is also heavily affected by dust attenuation. 
The energy of the UV photons is absorbed by the dust, and re-emitted in the IR. 
Therefore, there also exists a natural correlation between star formation rate and IR emission (see \citealt{Kennicutt12} for a review).

We here first consider the contribution from the obscured star formation activity, as observed in the rest-frame IR range (SFR$_{\rm IR}$).
For J2100c, J0842c and PJ231c, we use the values obtained from the Arp 220 SED (see Section \ref{secAnalysisSED} and Table \ref{tabSFRMstar}).
In case of PJ167c, we follow the method described in Section \ref{secAnalysisSED} to derive its SFR$_{\rm IR}$, but, instead of Arp 220, we use the best SED from the \texttt{MAGPHYS}--highz fit (see Figure \ref{figSEDobs} and Table \ref{tabSFRMstar}).
An alternative way of computing the star formation rate is through the luminosity of the [\cii] emission line ($\rm L_{[\scii]}$; e.g.~\citealt{DeLooze11}, \citeyear{DeLooze14}, \citealt{Sargsyan12}, \citealt{HerreraCamus15}).
Here, we take the values of SFR$_{\rm [\scii]}$ reported in \citeauthor{Decarli17} (\citeyear{Decarli17}), ranging from $\sim$260 to $\sim$730 M$_{\odot}$ yr$^{-1}$, i.e.~of the same order of magnitude as those measured from the dust continuum.
For PJ167c, we consider the measurement of the [\cii] line from recent high-resolution ALMA observations, i.e.~$F_{\mathrm{[\cii]}}=1.24 \pm 0.09$ Jy km s$^{-1}$ (Neeleman et al.~2019). We measured the corresponding [\cii] luminosity and star formation rate following \cite{Carilli13} and \cite{DeLooze14}, respectively. 
In Table \ref{tabSFRMstar} we report all the SFR$_{\rm [\scii]}$ values.

On the other hand, we can obtain measurements of (or limits on) the un-obscured contribution to the SFR in the companions, using our \textit{HST}/WFC3 sensitive observations in the F140W filter.
We consider the conversion between far UV (0.155 $\mu$m) luminosity ($L_{\rm FUV}$) and SFR$_{\rm UV}$ provided by \cite{Kennicutt12}:
\begin{equation}
\mathrm{log} \left[ \frac{\rm SFR_{\rm UV}}{\rm M_{\odot}\ yr^{-1}} \right] =\mathrm{log} \left[ \frac{L_{\rm FUV}}{\rm erg\ s^{-1}} \right] -C_{\rm FUV} 
\end{equation}
with $C_{\rm FUV}$=43.35. We report in Table \ref{tabSFRMstar} the estimated SFR$_{\rm UV}$ values. 
The limits achieved by our data are very sensitive, down to few M$_{\odot}$ yr$^{-1}$.
PJ167c, the only companion detected in the rest--frame UV, has an inferred un-obscured star formation rate of $\sim$11 M$_{\odot}$ yr$^{-1}$.
We note that the central wavelength of the broad band filter used here (F140W) corresponds to $\lambda_{\rm rest}\sim$0.18--0.2 $\mu$m for $z\sim$6-6.6, i.e.~in between the classically defined FUV and near UV (NUV; 0.230 $\mu$m).
In order to check how this impacts our results, we repeat our star formation rate estimates considering the calibration for the NUV ($C_{\rm NUV}$=43.17; \citealt{Kennicutt12}).
In this case, we measure SFR values only $\sim1.5\times$ larger. 
We also consider the best SED fit from \texttt{MAGPHYS}--highz for PJ167c, and we calculate the star formation rate in the exact FUV range. We obtain SFR$_{\rm UV}\sim$8 M$_{\odot}$ yr$^{-1}$, consistent, within the errors, with the one measured directly from our \textit{HST} data.
\begin{figure}[!h]
\centering
\includegraphics[width=\columnwidth]{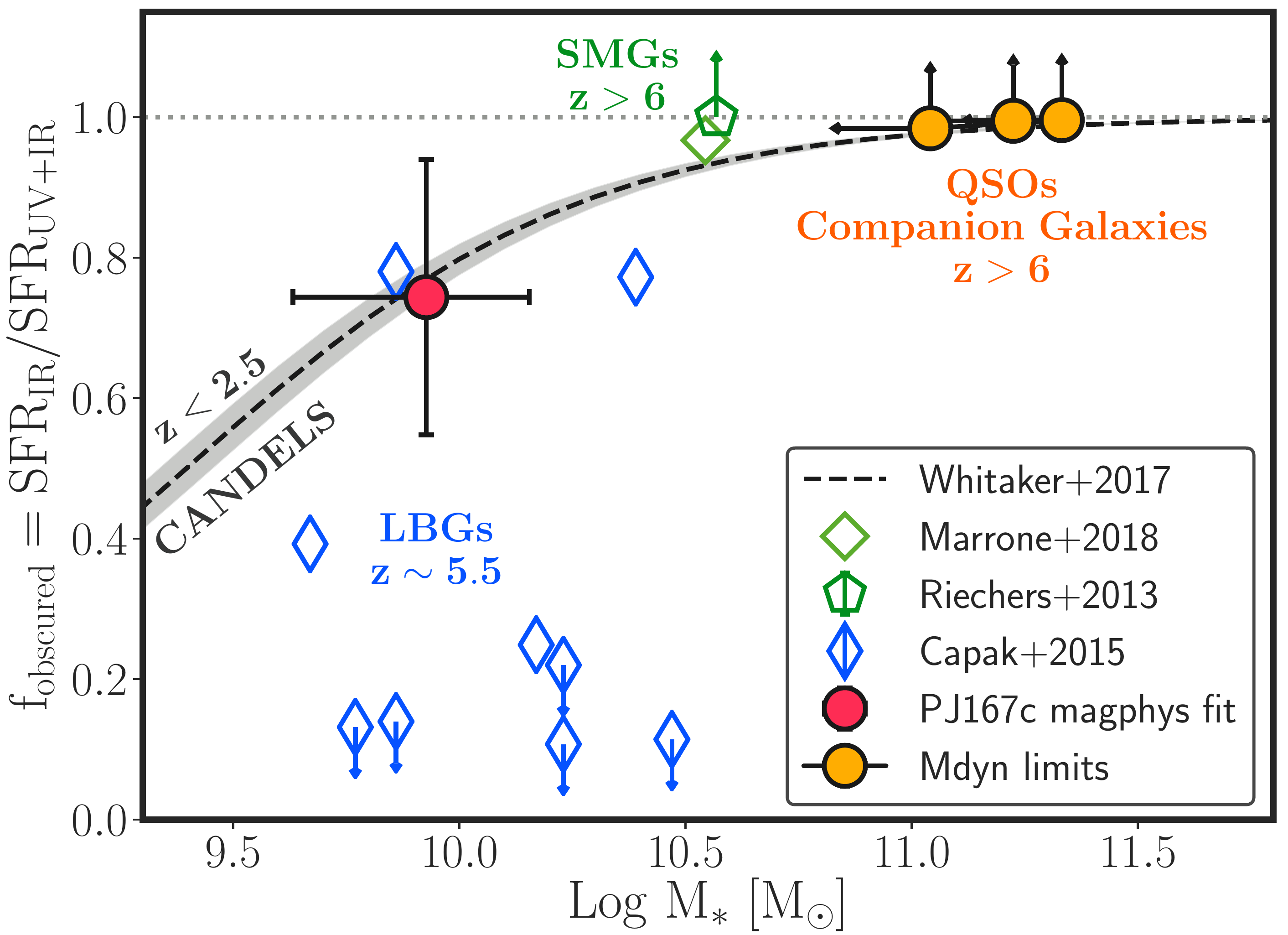}
\caption{
Fraction of obscured star formation as a function of stellar mass.
A tight correlation is observed at lower redshifts (0$<z<$2.5; \textit{dashed black line}, \citealt{Whitaker17}).
We show the location of $z>$6 SMGs observed by \citeauthor{Marrone18} (\citeyear{Marrone18}; \textit{big diamond}) and \citeauthor{Riechers13} (\citeyear{Riechers13}; \textit{pentagon}), together with $z\sim$5.5 LBGs from \citeauthor{Capak15} (\citeyear{Capak15}; \textit{blue diamond}).
The galaxies studied in this work are reported with \textit{red} (PJ167c, whose physical properties were obtained with the code \texttt{MAGPHYS}--highz) and \textit{yellow circles} (J2100c, PJ231c, J0842c; see Section \ref{secAnalysisSED}). In the latter case, only limits for the un--obscured SFR could be derived (see Section \ref{secAnalysisSFRUV}). 
The star formation of companions to high--$z$ quasars is dominated by the obscured component.
}
\label{figMstarFobs}
\end{figure}

With the exception of PJ167c, the SFRs measured in the IR in the companions studied here are $\sim$two orders of magnitude larger than the limits we set for the companions rest-frame UV emission.
The contribution of SFR$_{\rm UV}$ to the total star formation budget is therefore negligible.
In case of PJ167c, the un--obscured star formation rate is instead only $\sim 6\times$ lower than the obscured one.
Another way of performing this comparison is by looking at the fraction of obscured star formation, defined as $f_{\rm obscured}={\rm SFR}_{\rm IR}/{\rm SFR}_{\rm IR+UV}$.
\cite{Whitaker17} reported a tight correlation between this quantity and the stellar mass, irrespective of redshift (up to $z<$2.5), in a large sample of star--forming galaxies from CANDELS and SDSS.
We calculate (limits on) $f_{\rm obscured}$ for the galaxies presented here. We report these values in Table \ref{tabSFRMstar}, and we show them in the context of previous observations in Figure \ref{figMstarFobs}.
Again, the star formation rate of the companions is highly dominated by SFR$_{\rm IR}$, with obscured fractions ranging between 0.74 and 0.99.
In particular, taking into account the large uncertainties on $M_{*}$ and $f_{\rm obscured}$, PJ167c is consistent with the expectations from lower$-$redshift studies.
The remaining sources seem to also follow the $z<$2.5 trend (with the caution that we are here only able to set upper limits on their stellar masses).
\subsection{\textit{SFR vs Stellar Mass}} \label{secAnalysisMSFR}
A large number of studies has found a correlation between the SFR and the stellar mass of star--forming galaxies (``main sequence'', MS) over a wide redshift range ($0 \lesssim z \lesssim 6$; e.g.~\citealt{Brinchmann04}, \citealt{Noeske07}, \citealt{Whitaker11}, \citealt{Rodighiero11}; for an in--depth analysis of the literature see \citealt{Speagle14}).
The tightness ($\sim$0.3--0.2 dex scatter) of this relation has been interpreted as evidence that ``regular'' star--forming galaxies have smooth star formation histories, in which the majority of the mass is assembled via steady accretion of cool gas from the intergalactic medium on long timescales (e.g.~\citealt{Daddi07}, \citealt{Steinhardt14}).
On the other hand, highly starbursting galaxies, which lie above the MS, are also observed, and are thought to grow mainly via efficient, merger--triggered star formation events (e.g.~\citealt{Santini14}).
The MS normalization is observed to evolve with redshift, and this trend suggests that higher specific star formation rates are common at early cosmic times (e.g.~\citealt{Whitaker14}). 

We compare the properties of the companion galaxies considered here with those of typical star forming galaxies and SMGs at similar redshifts (see Figure \ref{figSFRMstarObs}).
We consider the observed MS relation at $z\sim6$ provided by \cite{Salmon15} and \cite{Speagle14}, together with predictions from semi-analytical models by \citeauthor{Somerville08} (\citeyear{Somerville08}, \citeyear{Somerville12}).  
\cite{Salmon15} examine $3.5 \leq z \leq 6.5$ galaxies in the GOODS-S field:
\begin{figure}[!h]
\centering
\includegraphics[width=\columnwidth]{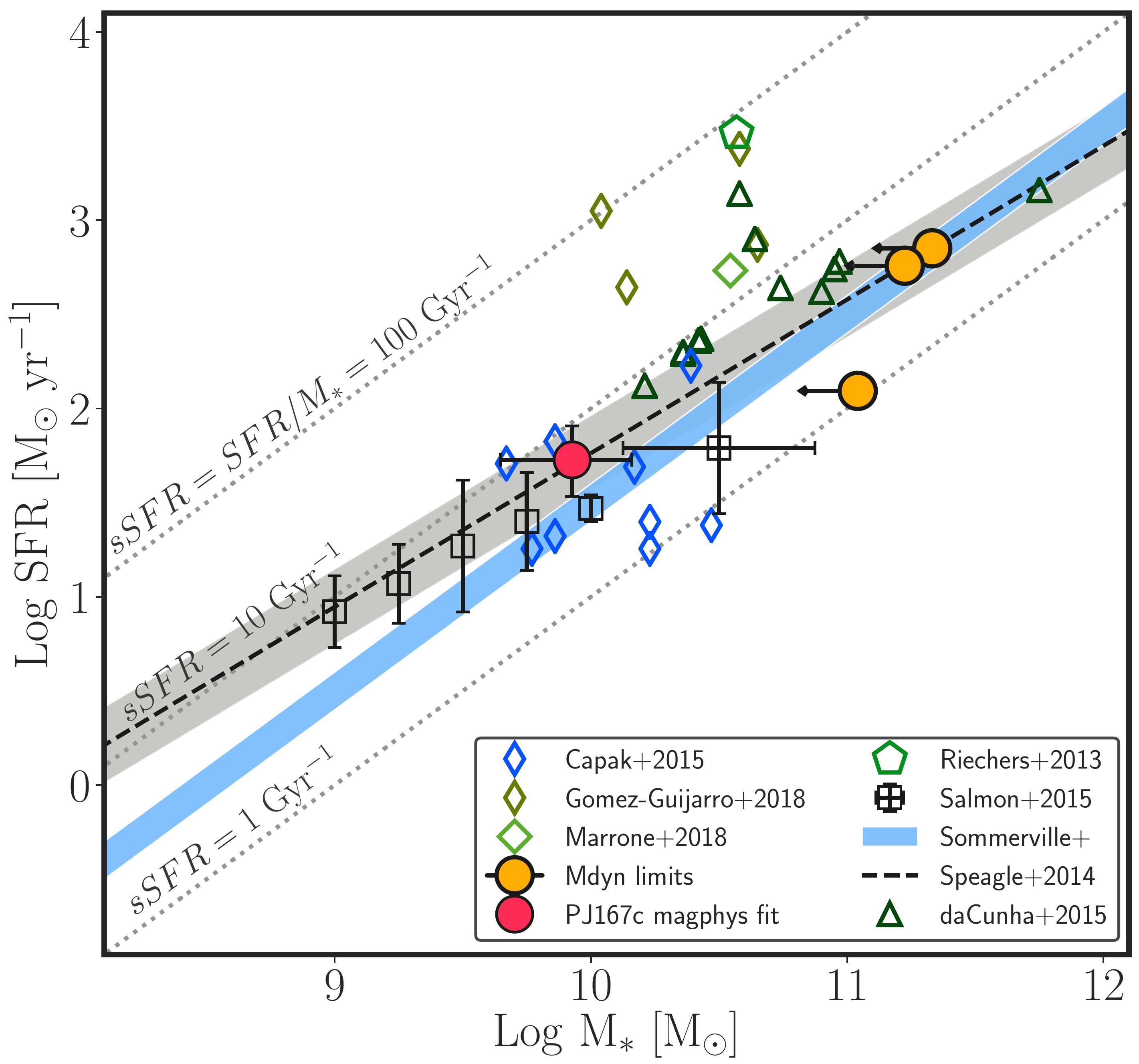}
\caption{
Star formation rate as a function of stellar mass for a compilation of sources at $z\sim$6.
We report observations of the galaxy main sequence (MS) from \citeauthor{Salmon15} (\citeyear{Salmon15}; \textit{empty black squares}), the empirically derived MS relation by \citeauthor{Speagle14} (\citeyear{Speagle14}; \textit{dashed line and grey region}), and the MS location predicted by semi-analytical models  (\citealt{Somerville12}; \textit{light blue region}).
The Speagle et al.~relation is based on observations with $M_{*}<10^{10.5}$ M$_{\odot}$, and extrapolated linearly at higher masses.
We show further examples of sub-millimeter galaxies, from $z\sim4.5-6.1$ sources (\citealt{DaCunha15}, \textit{triangles}, and \citealt{Gomez18}, \textit{small diamonds}) to the extreme starbursts observed at $z=$6.3 (\citealt{Riechers13}; \textit{pentagon}) and at $z=$6.9 (\citealt{Marrone18}; \textit{big diamond}), and $z \sim$5.5 LBGs from \citeauthor{Capak15} (\citeyear{Capak15}, \textit{blue diamonds}). We note that the SFR of the galaxies taken from the literature are derived with different methods (see Section \ref{secAnalysisMSFR}).
The companion galaxies reported in this work are shown with \textit{red} and \textit{yellow circles} (labels analogous to Figure \ref{figMstarFobs}).
Finally, we show the loci of constant sSFRs (\textit{gray dotted lines}). 
The companion galaxies are consistent with lying on the MS at $z\sim$6.
Deeper observations, particularly in the rest-frame optical region, are necessary to securely characterize the properties of these sources.
}
\label{figSFRMstarObs}
\end{figure}
we take here SFR and $M_{*}$ values of their $\sim$200 $z \sim 6$ galaxies.
\cite{Speagle14} assemble a comprehensive compilation of 25 studies of the MS at $ 0 \lesssim z\lesssim$6.
After a careful recalibration of the various datasets,
they obtain a robust SFR$-M_{*}$ relation as a function of the age of the universe ($t$, here in Gyr):  
\begin{equation}
\mathrm{logSFR} [M_{*},t] = (0.84 - 0.026 \times t) \mathrm{log} M_{*}-(6.51 - 0.11 \times t)
\end{equation}
They also find a scatter around the MS of $\sim$0.2 dex, irrespective of redshift.
We show this relation, calculated at $z=6$ with the representative 0.2 dex scatter, in Figure \ref{figSFRMstarObs}.
We consider the semi-analytical model by \cite{Somerville12}, who use N-body simulations and several feedback/accretion recipes to specifically reproduce the GOODS-S field. In particular, we consider the MS relation for this model at $z\sim$6, as provided by \citeauthor{Salmon15} (\citeyear{Salmon15}; see their Table 4).
In addition, we report observed SMGs at $4.5 < z < 6.1$ from \cite{DaCunha15}, whose redshifts and physical parameters were obtained with \texttt{MAGPHYS}-highz, and at $z \sim$4.5 from \cite{Gomez18}, for which recent ALMA mm observations and secure spectroscopic redshifts are available.
Finally, we show the massive, extremely starbursting galaxies at $z>6$ discovered by \cite{Riechers13} and \citeauthor{Marrone18} (\citeyear{Marrone18}; see Section \ref{secIntro}), and $z\sim$5.5 LBGs \citep{Capak15}.

We show in Figure \ref{figSFRMstarObs} the SFR and $M_{*}$ values obtained with \texttt{MAGPHYS}--highz for PJ167c. This galaxy lies on the MS at $z\sim$6.
For the remaining galaxies, i.e.~J2100c, J0842c and PJ231c, we only consider the obscured star formation rates and the upper limits on the stellar masses (see Section \ref{secAnalysisSED}).
These highly conservative constraints place the companions on or above the MS relation.

Future, deeper observations in the IR regime, together with further development of current fitting machines 
,will be needed to constrain these galaxies SEDs and stellar masses.
\section{Conclusions} \label{secConc}
In this work, we present sensitive follow-up optical and NIR imaging and spectroscopy of companion galaxies adjacent (i.e.~$< 60$ kpc and $<$450 km s$^{-1}$) to four $z\sim 6$ quasars, initially discovered by their bright [\cii] and far--infrared emission with ALMA (\citealt{Decarli17}, \citealt{Willott17}).

The data reported here have been acquired with several ground- and space--based facilities (i.e.~VLT/MUSE, MODS and LUCI at the LBT, Magellan/FIRE, \textit{Spitzer}/IRAC and \textit{HST}/WFC3),
and are aimed at probing the galaxies stellar content, recovered in the rest--frame UV/optical regime.
We perform aperture photometry at the location of the galaxies (as measured by ALMA), after accounting for both the bright, point--like, non--thermal quasar radiation and any foreground objects.
We detect no rest--frame 5000-7000 \AA~stellar emission (at $>3\sigma$ significance level) from the companions, observed at 3-5 $\mu$m.
In addition, no light from young stars, probed at $\lambda_{\mathrm{obs}} \sim$1.4 $\mu$m by \textit{HST}/WFC3, is detected in three of the four sources examined, i.e.~J2100c, J0842c and PJ231c.
However, the companion galaxy of the quasar PJ167 is detected in our \textit{HST} observations at 6.4$\sigma$.

From a comparison with SEDs of various local galaxies, we find that the companions PJ231c, J2100c and J0842c are consistent with an Arp 220--like galaxy at $z\sim6$.
These objects are heavily dust--obscured and/or they harbor a modest stellar mass.
The source PJ167c resembles, instead, a less extreme star--forming galaxy.
We compute SFRs and $M_{*}$ with the SED fitting code \texttt{MAGPHYS}--highz for this galaxy, whose emission is detected in more than one broad band.
We derive the obscured SFR of PJ231c, J0842c and J2100c by assuming the SED of Arp 220 scaled at the observed fluxes.
We place upper limits on their stellar masses by subtracting their  gas masses, estimated from the dust content, from their total dynamical masses, derived from the [\cii]~emission line widths.
We also derive tight constraints on their un-obscured star formation rate, as obtained from the sensitive $HST$/WFC3 data. We observe SFR$_{\rm FUV}\lesssim$3 M$_{\odot}$ yr$^{-1}$, i.e.~more than two orders of magnitude lower than SFR$_{\rm IR}$, with the exception of PJ167c, whose obscured star formation component is only $\sim$6$\times$ larger than the un-obscured value.
Finally, we find that the companions examined here are consistent with being on the main sequence of star forming galaxies at $z\sim6$.
However, our constraints/limits, in particular on the stellar masses, are still coarse.
This is mainly due to the lack of detections in the bluer bands.

In the near future, deep observations with upcoming instruments, e.g.~the NIRCAM and NIRSPEC cameras on board the \textit{James Webb Space Telescope}, will enable us to uncover the emission and dynamics of the stellar content of these galaxies, and, together with updated fitting techniques, to place strong constraints on their SEDs.

\vspace{20pt}

We are grateful to the anonymous referee for constructive feedback.
We thank J.~Heidt for the acquisition of the LBT/LUCI data.
CM acknowledges G.~Popping, A.~Drake, N.~Kacharov and E.~Da Cunha for useful insights on galaxy spectral modeling, and for support on the use of \texttt{MAGPHYS}.
CM thanks I.~Georgiev, K.~Jahnke and A.~Merritt for precious advice on PSF modeling and subtraction. 

BPV and FW acknowledge funding through the ERC grants ``Cosmic Dawn'' and ``Cosmic Gas''.
DR acknowledges support from the National Science Foundation under grant number AST-1614213.
CM thanks the IMPRS for Astronomy and Cosmic Physics at the University of Heidelberg. 

The present work is based on observations taken with ESO Telescopes at the La Silla Paranal Observatory, under the programs: 099.A-0682, 297.A-5054

This paper includes data gathered with the 6.5 meter Magellan Telescope located at Las Campanas Observatory, Chile.

The LBT is an international collaboration among institutions in the United States, Italy and Germany. LBT Corporation partners are: The University of Arizona on behalf of the Arizona university system; Istituto Nazionale di Astrofisica, Italy; LBT Beteiligungsgesellschaft, Germany, representing the Max-Planck Society, the Astrophysical Institute Potsdam, and Heidelberg University; The Ohio State University, and The Research Corporation, on behalf of The University of Notre Dame, University of Minnesota and University of Virginia.
This paper used data obtained with the MODS spectrographs built with
funding from NSF grant AST-9987045 and the NSF Telescope System
Instrumentation Program (TSIP), with additional funds from the Ohio
Board of Regents and the Ohio State University Office of Research.

Based on observations made with the NASA/ESA Hubble Space Telescope, obtained from the Data Archive at the Space Telescope Science Institute, which is operated by the Association of Universities for Research in Astronomy, Inc., under NASA contract NAS 5-26555. These observations are associated with program 14876. Support for this work was provided by NASA through grant number 10747 from the Space Telescope Science Institute, which is operated by AURA, Inc., under NASA contract NAS 5-26555. 

This work is based [in part] on observations made and archival data obtained with the Spitzer Space Telescope, which is operated by the Jet Propulsion Laboratory, California Institute of Technology under a contract with NASA. Support for this work was provided by NASA through an award issued by JPL/Caltech.

This work has made use of data from the European Space Agency (ESA)
mission {\it Gaia} (\url{https://www.cosmos.esa.int/gaia}), processed by
the {\it Gaia} Data Processing and Analysis Consortium (DPAC,
\url{https://www.cosmos.esa.int/web/gaia/dpac/consortium}). Funding
for the DPAC has been provided by national institutions, in particular
the institutions participating in the {\it Gaia} Multilateral Agreement.

This research made use of Astropy, a community-developed core Python package for Astronomy (Astropy Collaboration, 2018;  http://www.astropy.org).

\textit{Facilities:} VLT:Yepun (MUSE), Magellan:Baade (FIRE), LBT (MODS, LUCI), ALMA, \textit{HST} (WFC3), \textit{Spitzer} (IRAC).
\appendix
\section{A. A dust-continuum emitting source adjacent to the quasar VIK J2211$-$3206} \label{appJ2211}
We detect emission from the dust continuum, but not from the [\cii] emission line, from a source in the field of the quasar J2211 (QSO R.A.~22:11:12.39 ; Decl.~-32:06:12.9), at redshift $z_{\rm quasar} =$6.3394 $\pm$ 0.001 (\citealt{Decarli18}).
No secure redshift value is measured for this neighboring source (hereafter J2211c).
Note that the detection of an object with flux density comparable to J2211c over the area covered in the ALMA survey \citep{Decarli18} is expected from a comparison with the number counts of 1.2mm--bright sources observed in blank fields (e.g.~\citealt{Aravena16}).
Indeed, if one integrates the luminosity function of 1.2mm detected--sources provided by \cite{Fujimoto16} down to the flux of J2211c (see Table \ref{tabJ2211}), one expects $\sim$2.4 sources in 1 arcmin$^{2}$.
This amounts to $\sim$9.8 sources in the effective area spanned by our ALMA Survey (i.e.~$\sim$4 arcmin$^{2}$). This number is consistent with that of sources with similar brightness as J2211c (10) found in the sample recently compiled by \cite{Champagne2018}.

\begin{figure*}[!h]
\centering
\includegraphics[width=0.7\textwidth]{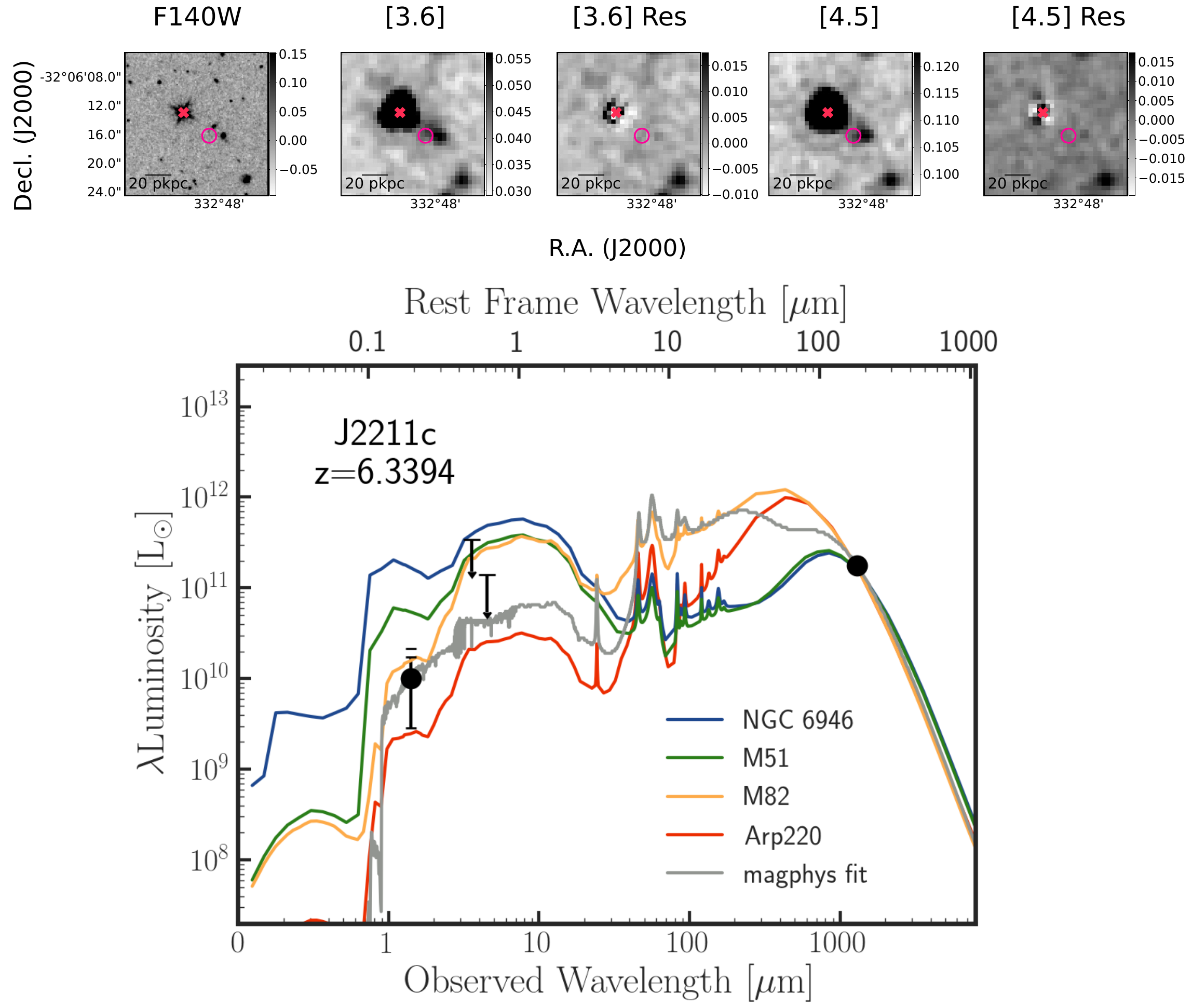}
\caption{Source adjacent to the quasar J2211, detected solely in the dust--continuum emission, i.e.~with no secure redshift measure.
\textit{Top}: Postage stamps (20\arcsec$\times$20\arcsec) of our follow--up observations; labels are as in Figure \ref{figPSall}.
\textit{Bottom}: Spectral Energy Distribution of J2211c.
We assume that the source is located at the same redshift of the quasar.
We report our photometric measurements/limits and, for comparison, various templates of local galaxies and the best SED fit from \texttt{MAGPHYS}-highz. The labels and templates are as in Figure \ref{figSEDobs}. 
J2211c SED results to be intermediate between the low--$z$ ULIRG Arp220 and the starbursting galaxy M82 (see Section \ref{secAnalysisSED}). 
On the basis of our follow--up observations, and considering the predicted density of mm--sources, we are not able to exclude that this source is a fore/background  (see text for details, and \citealt{Champagne2018}).
}
\label{figJ2211}
\end{figure*}

We acquired new observations of this field as part of our follow--up campaign of [\cii]--bright companions to high--redshift quasars, using $HST$/WFC3 and $Spitzer$/IRAC (see Table \ref{tabObserv} for details of the observations). 
We reduced and analyzed the data following the procedures reported in Section \ref{secObs}.
In what follows, we assume that J2211c is located at the redshift of the quasar. 
No emission from the stellar population in the rest--frame optical regime is measured (at 3$\sigma$ significance) in the \textit{Spitzer}/IRAC images. 
However, we tentatively measure ($S/N=$2.1) emission in the F140W filter with the \textit{HST}/WFC3 camera.
We report our photometric measurements/3$\sigma$ limits in Table \ref{tabJ2211}, where we also list the galactic properties (coordinates and mm flux) obtained from ALMA data (\citealt{Decarli17}, \citealt{Champagne2018}).
In Figure \ref{figJ2211} we show the postage stamps of our follow--up observations.

In analogy with the companions discussed in the main body of the paper, we compare the spectral energy distribution of J2211c with those of local galaxies, and we fit our photometric data with \texttt{MAGPHYS}--highz (see Figure \ref{figJ2211}).
From the latter, we find that the SED of J2211c is better reproduced by a galaxy model in between Arp 220 and M82 (i.e.~a powerful local ULIRG and a starburst), with $M_{*}\sim3 \times 10^{10}$ M$_{\odot}$ and SFR$\sim$130 $M_{\odot}$ yr$^{-1}$.
We further measure the obscured/un--obscured SFR ratio of J2211c, following the procedure used for PJ167c (see Section \ref{secAnalysisSFRUV}).
The star formation rate is dominated by the obscured contribution (SFR$_{\rm UV}\sim$2 $M_{\odot}$ yr$^{-1}$ and $f_{\rm obscured}\sim$0.99).
We report all these estimates in Table \ref{tabJ2211}.
The lack of a secure redshift confirmation prevents us from drawing further conclusions on the nature of this source, or from placing it in the context of previous observations.
\begin{deluxetable*}{lc}[!h]
\tabletypesize{\small}
\tablecaption{Information on VIK J2211$-$3206c, a source adjacent to the quasar J2211 detected only via its dust continuum emission.
Given the lack of any redshift measurement, we are not able to securely identify this galaxy as physically interacting with the quasar, and place it in the context of the analysis the companions.
We report here its coordinates and projected spatial separation to the quasar, obtained from ALMA data (\citealt{Decarli18}, \citealt{Champagne2018}), and our $HST$/WFC3 and $Spitzer$/IRAC follow--up photometric measurements/limits (see Figure \ref{figJ2211}).
We also list our constraints on its physical properties, given the assumption that J2211c lies at the quasar redshift.
\label{tabJ2211}}
\tablewidth{0pt}
\tablehead{ \colhead{\qquad \qquad \qquad \qquad VIK J2211$-$3206c}}
\startdata
R.A.~(J2000) &  22:11:12.11  \\
Decl.~(J2000) & -32:06:16.19\\
$\Delta$r$_{\mathrm{projected}}$ [kpc] & 26.8 \\
F140W [mag] & 27.39 $\pm$ 0.52 \\
$F_{\mathrm{3.6  \mu m}}$ [$\mu$Jy] & $<$3.42  \\
$F_{\mathrm{4.5  \mu m}}$ [$\mu$Jy] & $<$1.80\\
$F_{\mathrm{mm}}$ [mJy] & 0.64 $\pm$ 0.06 \\
SFR$_{\rm IR}$  [M$_{\odot}$ yr$^{-1}$] &  257 $\pm$ 36 \\
SFR$_{\rm UV}$  [M$_{\odot}$ yr$^{-1}$] & 2 $\pm$ 2  \\
$f_{\rm obscured}$ & 0.99 $\pm$ 0.14 \\
SFR$_{\mathrm{magphys}}$  [M$_{\odot}$ yr$^{-1}$] & 132$^{+120}_{-59}$ \\
M$\rm _{*, magphys}$ [M$_{\odot}$] & 2.75$^{+3.13}_{-1.47} \times$ 10$^{10}$ 
\enddata
\end{deluxetable*}
\section{B. Quasar photometry} \label{appQSO}
\begin{figure*}[!h]
\centering
\includegraphics[width=\textwidth]{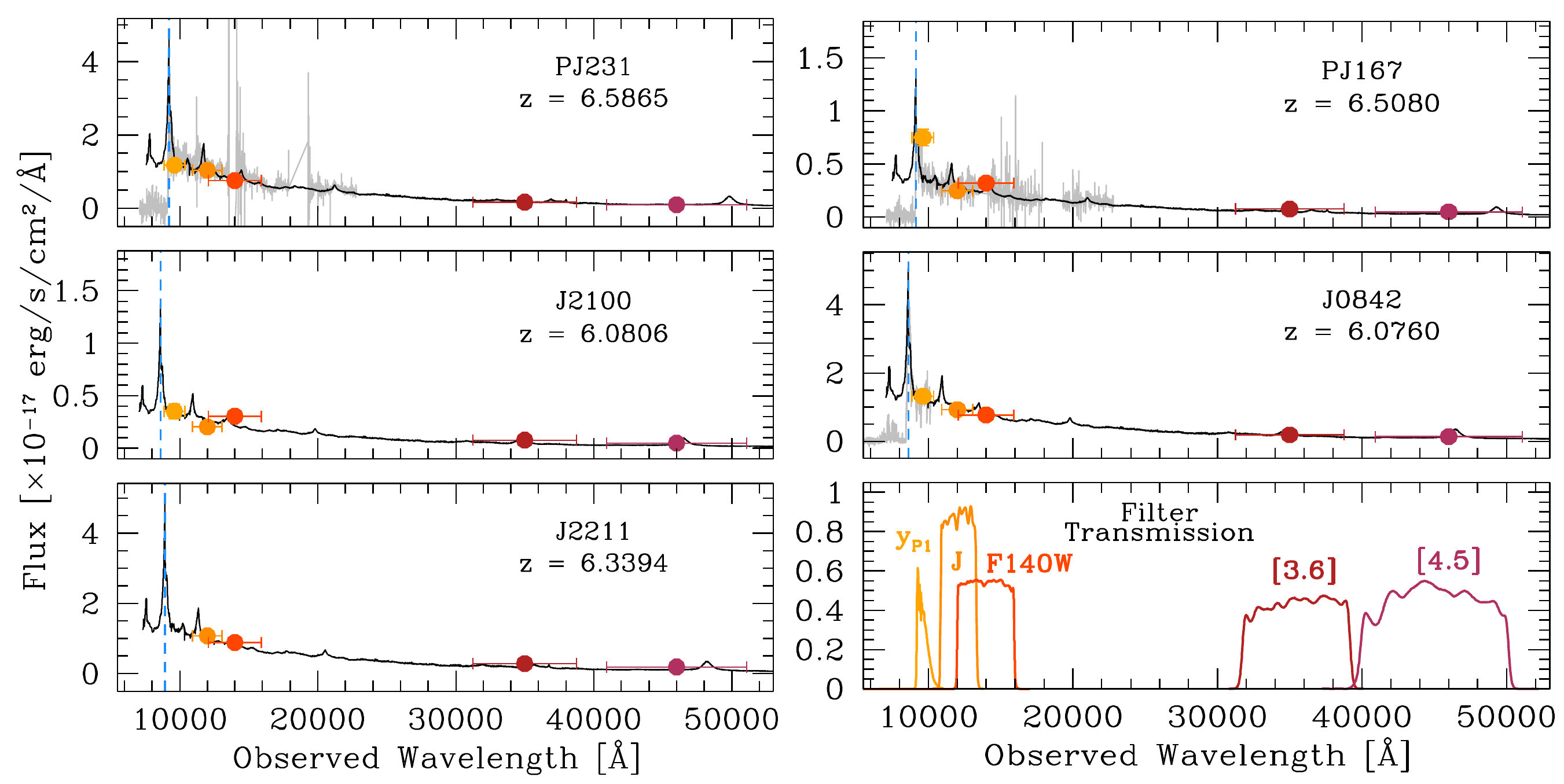}
\caption{
Spectral Energy Distribution of the quasars in our sample. The observed photometric measurements (\textit{filled points}) are obtained from our new follow-up data and from the literature (see Table \ref{tabQSOPhot}; the filter responses are reported in the \textit{lower right panel}). We also show the available optical/NIR spectra (\textit{light grey}; see also \citealt{Mazzucchelli17b}), and a lower-redshift composite template shifted at the redshift of the quasar (\textit{black solid line}; \citealt{Selsing16}). The location of the Ly$\alpha$ line is highlighted with a \textit{light blue dashed line}.
}
\label{figappQSOphot}
\end{figure*}

In the framework of our study of companion galaxies, we also perform forced photometry at the position of the quasars in the \textit{Spitzer}/IRAC and \textit{HST}/WFC3 images (see Section 2 for methodology).
In Table \ref{tabQSOPhot} we report the derived quasars' photometry.
In Figure \ref{figappQSOphot}, we show the quasars SEDs. The fluxes measured in our follow-up data are consistent with those expected from the observed optical/NIR spectra, when available, and/or from a lower$-z$ quasar template \citep{Selsing16} re-scaled to match the observed $J$ band magnitude. 
\begin{deluxetable*}{lccccccccc}[!h]
\tabletypesize{\small}
\tablecaption{Photometric measurements of the quasars studied in this work (see Section \ref{secObs}). The measurements in the y$_{P1}$ band are from the PS1 PV3 catalog, while the $J$ band values are from : (1) \cite{Jiang15}; (2) \cite{Venemans15b}; (3) \cite{Mazzucchelli17b}; (4) \cite{Willott10a}; (5) Venemans et al.~in prep..
\label{tabQSOPhot}}
\tablewidth{0pt}
\tablehead{
\colhead{name} & \colhead{$F_{\mathrm{y_{P1}}}$}  & \colhead{$F_{J}$} & \colhead{$F_{\mathrm{F140W}}$} & \colhead{$F_{\mathrm{3.6  \mu m}}$} & \colhead{$F_{\mathrm{4.5  \mu m}}$} &
\colhead{$F_{\mathrm{1.2 mm}}$} &
\colhead{Ref $J$}
\\
 & [mJy] & [mJy] & [mJy] & [mJy] &  [mJy] & [mJy] 
}
\startdata
SDSS J0842+1218 & 40.55$^{+2.3}_{-2.2}$ & 44.46$\pm$1.2 &  50.19$\pm$0.02 & 76.29$\pm$0.22 & 93.33$\pm$0.25 & 0.87$\pm$0.18 & (1) \\ 
PSO J167.6415--13.4960 & 23.12$^{+2.5}_{-2.2}$ & 11.91$\pm$1.0  & 20.89$\pm$0.02 & 30.56$\pm$0.23 & 34.32$\pm$0.17 & 0.87$\pm$0.05 & (2) \\
PSO J231.6576--20.8335 & 36.31$^{+2.8}_{-2.6}$ & 49.66$^{+2.3}_{-2.2}$ & 49.13$\pm$0.02 & 66.95$\pm$0.22 & 67.91$\pm$0.22 & 4.41$\pm$0.16 & (3) \\
CFHQS J2100$-$1715 & 10.86$^{+2.3}_{-1.9}$ & 9.82$\pm$0.9 & 19.95$\pm$0.02 & 31.65$\pm$0.25 & 34.76$\pm$0.26 & 1.20$\pm$0.15 & (4) \\
SDSS J2211--3206 & -- & 51.52 $_{-4.5}^{+5.0}$ & 57.93$\pm$0.02 &  116.05$\pm$0.23 & 131.45$\pm$0.19 & 0.57$\pm$0.05 & (5)
\enddata
\end{deluxetable*}
\vspace{2mm}

\clearpage

\clearpage

\end{document}